\documentclass[preprint,12pt]{elsarticle}

\usepackage{amssymb}
\usepackage{amsmath}
\usepackage{braket}

\journal{Annals of Physics}

\begin{document}

\begin{frontmatter}

\title{A new approach for open quantum systems \\
based on a phonon number representation of \\
a harmonic oscillator bath}

\author[A1]{M. Tokieda}\corref{cor1}
\author[A1,A2]{K. Hagino}
\address[A1]{Department of Physics, Tohoku University, Sendai 980-8578,  Japan}
\address[A2]{Research Center for Electron Photon Science, Tohoku University, 1-2-1 Mikamine, Sendai 982-0826, Japan}
\cortext[cor1]{Corresponding author. \\
{\it E-mail addresses:} \\
tokieda@nucl.phys.tohoku.ac.jp (M. Tokieda), \\
hagino@nucl.phys.tohoku.ac.jp (K. Hagino). \\
{\it Full postal address:} \\
6-3, Aoba, Aramaki, Aobaku, SENDAI, MIYAGI 980-8578, Japan
}

\begin{abstract}

To investigate a system coupled to a harmonic oscillator bath, we propose a new approach based on a phonon number representation of the bath.
Compared to the method of the hierarchical equations of motion, the new approach is computationally much less expensive in a sense that a reduced density matrix is obtained by calculating the time evolution of vectors, instead of matrices, which enables one to deal with large dimensional systems. 
As a benchmark test, we consider a quantum damped harmonic oscillator, and show that the exact results can be well reproduced.
In addition to the reduced density matrix, our approach also provides a link to the total wave function by introducing new boson operators.

\end{abstract}


\end{frontmatter}


\section{Introduction}
\label{intro}

Open quantum systems have attracted enormous attention not only in physics, but also in science in general.
In those systems, the total Hamiltonian is composed of a Hamiltonian for a system of interest, that for an environment surrounding the system, and a coupling between the system and the environment \cite{UW08,BP02}.
When simulating an environment with a large number of degrees of freedom, the environment is often treated as a bath of harmonic oscillators. 
Applications of such treatment have ranged from nuclear physics \cite{MS80} to quantum biology \cite{IF12}.
In addition to its simplicity, one of the good reasons to employ a harmonic oscillator bath, especially in condensed matter physics and in nuclear physics, is that it leads to a Langevin equation in the classical limit \cite{UW08,CL81,FLO88}.
This fact enables one to discuss a quantum extension of the Langevin equation. 
Moreover, it should be noted that, as long as one considers only the reduced density matrix, one can find an equivalent harmonic oscillator bath to any kind of environment coupling to a system, when the coupling is not so strong \cite{MS80,FV63}.

For numerical studies of a harmonic oscillator bath, many methods have been developed so far. 
For instance, a method based on stochastic simulation has been considered in Refs.$\,$\cite{Breuer04,Stockburger04,Lacroix08,HS17}.
Although the formalism itself is robust, there are several drawbacks in this approach in practice.
One of the most harmful ones is that a calculation of each sample becomes unstable in nonlinear systems \cite{HL10}, which limits the applicability of the method.

Another powerful approach is the multilayer multiconfiguration time-dependent Hartree (ML-MCTDH) method \cite{WT08}.
In this method, one can obtain not only a reduced density matrix, but also the total wave function. 
However, one needs to take into account a large number of bath degrees of freedom, and, to our knowledge, an application of this method has been limited only to a spin-$1/2$ system, which is the simplest case in terms of numerics.

Until now, the most successful method in this regard is the hierarchical equations of motion (HEOM) \cite{TK89,YYLS04,IT05,Tanimura06,YWHXZY16}.
It introduces auxiliary density operators, and their equations of motion are constructed hierarchically. 
The method can deal with not only a system in the energy eigenstate representation, but also that in the coordinate space representation. 
Recently, the method has been extended to the imaginary time evolution, which represents an inverse temperature, and it is now possible to extract thermodynamic quantities based on this method \cite{Tanimura14,Tanimura15}.
It should be noted that the HEOM approach can be derived from a stochastic approach. 
In recent years, it is often implemented to formulate problems based on a stochastic approach, and then transform it into the HEOM to carry out numerical calculation \cite{YYLS04,Wu18,HC18}.

Since the HEOM method follows a time evolution of matrices which have the same dimension as a system under consideration, the numerical cost becomes expensive when dealing with a large dimensional system. 
To our knowledge, the largest system analyzed with the HEOM is a quantum spin glass with 12 spins on triangular lattices, for which the dimension is 4096 \cite{TT15}.
Although this was an important achievement, it has yet been demanded to develop a method which is not so sensitive to the dimension of a system. 
One step forward was made in Ref.$\,$\cite{NT18}, in which each element of the reduced density matrix was calculated based on wave functions.
In other words, when the dimension of a system is $N$, the calculation requires to follow only a time evolution of $N$-dimensional vectors, rather than that of $N \times N$ matrices. This reduction is especially important when $N$ is large. 
We mention that such reduction for the stochastic approach had been discussed by several authors \cite{Lacroix08,KZ16}.

In this paper, we develop an alternative approach to Ref.$\,$\cite{NT18} to solve the HEOM for vectors.
It is based on an expansion of the reduced density matrix with basis vectors of the number of phonon in a harmonic oscillator bath. 
Compared to Ref.$\,$\cite{NT18}, our method includes a certain approximation. 
However, as we will show, a benchmark calculation indicates that the method can well reproduce exact results.
Moreover, our method can also extract information on how much the bath degrees of freedom is excited in the course of time evolution.

This work can also be regarded as a new perspective of the coupled-channels (or the close-coupling) method with enormous number of harmonic oscillators. 
In the coupled-channels method, one expands the total wave function with respect to eigenstates of the internal Hamiltonian (See, for instance, Ref.$\,$\cite{HT12}).
If the internal motion is described by a single harmonic oscillator, it corresponds to an expansion with the phonon number basis.
Empirically, a convergence is obtained at several phonons, not so huge number, with a physical parameter set.
The problem is thus quite easy with a single harmonic oscillator, but what about with hundreds of harmonic oscillators, or with infinite number of harmonic oscillators ? 
Obviously, a direct application of the conventional coupled-channels approach is not feasible. 
The method which we propose in this paper can deal with such problems.

We would like to emphasize that this paper is not only for numerics. 
Firstly, our method has a clear physical background, that is, a phonon number representation of the bath. 
This enables one to attach physical meaning of quantities appeared in the formalism. 
Moreover, inspired by the coupled-channels method, we are also able to give a link to the total wave function.
Boson operators are naturally introduced in the course of discussion, and they will provide a new picture of open quantum systems.

The paper is organized as follows. 
Sec.$\,$\ref{method} is devoted to the methodology.
We first derive all necessary equations, including the HEOM to be solved and the formula for the reduced density matrix. 
In Sec.$\,$\ref{inter}, we explain the reason why the method can be interpreted as a phonon number representation.
In Sec.$\,$\ref{num}, we give several discussions regarding practical applications.
We particularly apply the method to a quantum damped harmonic oscillator, and we show that the exact results can be well reproduced. 
In Sec.$\,$\ref{link}, we further investigate the method, and discuss a link to the total wave function. For a continuous bath, we show that boson operators are naturally introduced.
Finally, we present the conclusion and future perspectives in Sec.$\,$\ref{conc}.


\section{Methodology}
\label{method}

In this section, the methodology of the new method is presented. 
Basically, it is the Taylor expansion of a part of the influence functional. 
It enables one to expand the reduced density matrix with vectors. 
Following the procedure of the conventional HEOM approach, equations for a time evolution of those vectors are derived.

\subsection{Introduction and basic strategy}
\label{method-basic}

In this paper, we consider a system coupled to a harmonic oscillator bath, which is now often referred to as the Caldeira-Leggett model \cite{CL81,CL83}.
The total Hamiltonian reads
\begin{equation}
\label{CL Hamiltonian 0}
\begin{gathered}
H_{\rm tot} = H_S(q,p) + \sum_{i} \left( \frac{p_i^2}{2m_i} + \frac{1}{2} m_i \omega_i^2 x_i^2  - \frac{\hbar \omega_i}{2} \right) + h(q) \sum_i c_i x_i \\
\equiv H_S + H_B + H_I.
\end{gathered}
\end{equation}
The coordinate and the momentum for the system of interest are denoted by $q$ and $p$, respectively, and $H_S(q,p)$ is Hamiltonian for the system. 
The so called counter term may be included in $H_S$ (see Eq.$\,$(\ref{HS dho}) below).
$x_i$, $p_i$, $m_i$, and $\omega_i$ are the coordinate, momentum, mass, and frequency of the $i$-th oscillator, respectively. 
In the definition of $H_B$, we have subtracted the zero point energy. 
The third term on the right hand side of Eq.$\,$(\ref{CL Hamiltonian 0}) is the interaction Hamiltonian, where $c_i$ is the strength of the interaction with the $i$-th oscillator and $h(q)$ is the interaction form factor.
We have here assumed that the interaction is separable between the system and the environment degrees of freedom.

For later discussions, we introduce the creation and the annihilation operators,
\begin{equation}
\begin{gathered}
a_i^{\dagger} = \sqrt{\frac{m_i \omega_i}{2 \hbar}} \left( x_i - \frac{i p_i}{m_i \omega_i} \right), \\
a_i = \sqrt{\frac{m_i \omega_i}{2 \hbar}} \left(x_i + \frac{i p_i}{m_i \omega_i} \right),
\end{gathered}
\end{equation}
where the dagger denotes the hermitian conjugate. 
These operators satisfy the boson commutation relation, $[a_i, a_j] = 0$ and $[a_i, a_j^{\dagger}] = \delta_{i,j}$ with the Kronecker delta $\delta_{i,j}$, and we use the term phonon to call these quanta.
Using these operators, the total Hamiltonian Eq.$\,$(\ref{CL Hamiltonian 0}) reads
\begin{equation}
\label{CL Hamiltonian}
H_{\rm tot} = H_S(q,p) + \sum_{i} \hbar \omega_i a_i^{\dagger} a_i + h(q) \sum_i d_i (a_i + a_i^{\dagger}),
\end{equation}
with $d_i = c_i \sqrt{\hbar / 2 m_i \omega_i}$.

We define the reduced density matrix $\rho_S(t)$ by taking the trace of the total density matrix $\rho(t)$ with respect to the bath degrees of freedom, that is, $\rho_S(t) = {\rm Tr}_B \rho(t)$. 
When one is interested in the time evolution of the reduced density matrix $\rho_{S}(t)$, the path integral description is useful. 
Denoting the eigenstates of the coordinates as $\ket{q,\vec{x}} \equiv \ket{q} \ket{\vec{x}} = \ket{q} \ket{x_1} \ket{x_2} \dots$, we employ the following notation for the path integral for the time evolution operator,
\begin{equation}
\label{path notation}
\braket{q_a,\vec{x}_a| e^{-i H_{\rm tot} t / \hbar } |q_c,\vec{x}_c} = \int_{(q_c,0)}^{(q_a,t)} D[Q] \int_{(\vec{x}_c,0)}^{(\vec{x}_a,t)} D[\vec{X}] \, e^{i S_{\rm tot}[Q,\vec{X},t] / \hbar},
\end{equation}
with the plank constant $\hbar$.
Here, the paths are taken which satisfy $Q(0) = q_c$, $Q(t) = q_a$, $\vec{X}(0) = \vec{x}_c$, and $\vec{X}(t) = \vec{x}_a$.
In this equation, $S_{\rm tot}[Q,\vec{X},t]$ is the action for the total Hamiltonian Eq.$\,$(\ref{CL Hamiltonian}), 
and can be decomposed into each part as $S_{\rm tot} = S_S + S_B + S_I$ with the obvious notation.

Using the path integral formulation, the time evolution of the reduced density matrix can be written as
\begin{equation}
\label{rhoS time evolution}
\begin{gathered}
\rho_S(q_a,q_b,t) \equiv \braket{q_a|\rho_{S}(t) |q_b} \\
= \int dq_c \int dq_d \ \rho_S(q_c,q_d,t=0) \\
\times \int_{(q_c,0)}^{(q_a,t)} D[Q] \int_{(q_d,0)}^{(q_b,t)} D^{*}[Q^{\prime}] \, e^{i \left( S_S[Q,t] - S_S[Q^{\prime},t] \right) / \hbar} \, \mathcal{F}[Q,Q^{\prime},t].
\end{gathered}
\end{equation}
We have here assumed the factorized initial condition $\rho(t=0) = \rho_{S}(t=0) \otimes \rho_B$. 
As is often the case, we consider a situation in which the initial condition of the system is a pure state, that is,
\begin{equation}
\label{S initial}
\rho_S(q_c,q_d,t=0) = \varphi(q_c) \varphi^{*}(q_d),
\end{equation}
with an initial wave function of a system $\varphi$.

In Eq.$\,$(\ref{rhoS time evolution}), $\mathcal{F}[Q,Q^{\prime},t]$ is called the Feynman-Vernon's influence functional defined by \cite{FV63}
\begin{equation}
\label{influence functional}
\begin{gathered}
\mathcal{F}[Q,Q^{\prime},t] = \int d\vec{x_c} \int d\vec{x_d} \ \rho_B(\vec{x_c},\vec{x_d}) \int d\vec{x} \\
\times \int_{(\vec{x}_c,0)}^{(\vec{x},t)} D[\vec{X}] \, e^{i \left( S_B[\vec{X},t] + S_I[Q,\vec{X},t] \right) / \hbar} \int_{(\vec{x}_d,0)}^{(\vec{x},t)} D^{*}[\vec{X}^{\prime}] \, e^{- i \left( S_B[\vec{X}^{\prime},t] + S_I[Q^{\prime},\vec{X}^{\prime},t] \right) / \hbar}.
\end{gathered}
\end{equation}
When each mode $i$ in $S_B$ is independent as in the present Hamiltonian Eq.$\,$(\ref{CL Hamiltonian}), and if the initial condition for $\rho_B$ is given in a separable form $\rho_B = \prod_i \rho_{B,i}$, the influence functional is also separable,
\begin{equation}
\label{prod if}
\mathcal{F}[Q,Q^{\prime},t] = \prod_i \mathcal{F}_i[Q,Q^{\prime},t].
\end{equation}

Note that the influence functional is a functional of $Q$ and $Q^{\prime}$, 
that is, it depends on the paths $Q(s)$ and $Q^{\prime}(s)$ for $0 \leqq s \leqq t$. 
Sometimes, the path corresponding to $Q(s)$ ($Q^{\prime}(s)$) is called the forward (the backward) path \cite{KZ16}.

From Eq.$\,$(\ref{rhoS time evolution}), one can see that the whole bath dependence is contained in the influence functional $\mathcal{F}$, as was emphasized in Ref.$\,$\cite{FV63}.
Hence, the time dependence of the reduced density matrix of two systems coincides with each other if their influence functionals are the same.
We will make use of this property in Sec.$\,$\ref{inter-finite}.

Given that the collection of harmonic oscillators is thermally equilibrated at initial time, 
that is, $\rho_B = \exp(-\beta H_B)/{\rm Tr}_B \exp(-\beta H_B)$ with an inverse temperature $\beta$, 
the influence functional for the Hamiltonian Eq.$\,$(\ref{CL Hamiltonian}) takes the following form \cite{CL81,FV63},
\begin{equation}
\label{CL if}
\mathcal{F}[Q,Q^{\prime},t] = f[Q,t] f^{*}[Q^{\prime},t] g[Q,Q^{\prime},t],
\end{equation}
with
\begin{equation}
\label{f def}
f[Q,t] = \exp \left( - \frac{1}{\hbar} \int_0^t ds \int_0^{s} d\tau \, h(Q(s)) h(Q(\tau)) L(s-\tau) \right),
\end{equation}
and
\begin{equation}
\label{g def}
g[Q,Q^{\prime},t] = \exp \left( \frac{1}{\hbar} \int_0^t ds \int_0^t ds^{\prime} h(Q(s)) h(Q^{\prime}(s^{\prime})) L(s^{\prime}-s) \right).
\end{equation}

In these equations, $L(t)$ is defined by
\begin{equation}
\label{bath correlation}
L(t) = \int_0^{\infty} d\omega J(\omega) \left[ \coth\left(\frac{\beta \hbar \omega}{2}\right) \cos(\omega t) - i \sin(\omega t) \right],
\end{equation}
with the spectral density,
\begin{equation}
\label{J}
J(\omega) = \frac{1}{\hbar} \sum_i d_i^2 \delta(\omega-\omega_i).
\end{equation}
By considering $J(\omega)$ being a continuous function of $\omega$, one can treat a continuous bath model.
It is helpful to define the negative argument of $J(\omega)$ as $J(-\omega) \equiv - J(\omega)$.
Then, the definition of $L(t)$ becomes more compact,
\begin{equation}
\label{bath correlation compact}
L(t) = \int_{-\infty}^{\infty} d\omega \frac{J(\omega)}{1-e^{-\beta \hbar \omega}} e^{-i \omega t}.
\end{equation}

In writing Eq.$\,$(\ref{CL if}), we have explicitly expressed the dependence of the influence functional on the forward and the backward paths.
$f[Q,t]$ and $f^{*}[Q^{\prime},t]$ in Eq.$\,$(\ref{CL if}) depend only on the forward and the backward paths, respectively, while the $g[Q,Q^{\prime},t]$ term describes their entanglement.

In Ref.$\,$\cite{KZ16}, the forward and the backward paths are disentangled by performing the Hubbard-Stratonovich transformation including ${\rm Re} \ln {f[Q,t]}$, ${\rm Re} \ln {f^{*}[Q^{\prime},t]}$, and $\ln g[Q,Q^{\prime},t]$, where ${\rm Re}$ means the real part. 
The authors of Ref.$\,$\cite{KZ16} then succeeded in deriving the hierarchical stochastic Schr\"odinger equations.
As is expected from Eq.$\,$(\ref{rhoS time evolution}) together with the initial condition Eq.$\,$(\ref{S initial}), one of advantages of this procedure is that one can derive the time evolution of the reduced density matrix by calculating vectors rather than matrices (see the following subsections for details).
As has been discussed in Sec.$\,$\ref{intro}, this is important to construct a method which is applicable to large dimensional systems.

Our approach proposed in this paper is similar to that in Ref.$\,$\cite{KZ16} in a sense that the forward and the backward paths are disentangled and that the time evolution of vectors is followed.
In contrast to Ref.$\,$\cite{KZ16}, which uses the Hubbard-Stratonovich transformation, we expand $g[Q,Q^{\prime},t]$ in the Taylor series,
\begin{equation}
\label{g taylor first}
g[Q,Q^{\prime},t] = \sum_{n=0}^{\infty} \frac{1}{n!} \left\{ \frac{1}{\hbar} \int_0^t ds \int_0^t ds^{\prime} h(Q(s)) h(Q^{\prime}(s^{\prime})) L(s^{\prime}-s) \right\}^n.
\end{equation}
In the rest of this section, we will show how one can calculate the time evolution of the reduced density matrix Eq.$\,$(\ref{rhoS time evolution}) with this expansion.
In addition to the methodology, we will also discuss a physical interpretation of the method in the next section, that is, the Taylor expansion up to the $n$-th order corresponds to taking into account up to the $n$-phonon states of the bath degrees of freedom.

\subsection{Expansions of $L(t)$}
\label{method-bath}

Let us first focus on the 0-th order of the expansion in Eq.$\,$(\ref{g taylor first}). 
In this simplest case, $g[Q,Q^{\prime},t] = 1$ and thus the influence functional is separable with respect to the forward and the backward paths,
\begin{equation}
\label{CL if 0}
\mathcal{F}[Q,Q^{\prime},t] = f[Q,t] f^{*}[Q^{\prime},t].
\end{equation}
Within this approximation, the time evolution of the reduced density matrix Eq.$\,$(\ref{rhoS time evolution}) takes a simple form,
\begin{equation}
\label{rhoS time evolution 0}
\rho_S(q_a,q_b,t) = \psi_0(q_a,t) \psi_0^{*}(q_b,t),
\end{equation}
with
\begin{equation}
\label{psi time evolution 0}
\psi_0(q_a,t) = \int dq_c \ \varphi(q_c) \int_{(q_c,0)}^{(q_a,t)} D[Q] \, e^{i S_S[Q,t] / \hbar } f[Q,t].
\end{equation}
One can follow the time evolution of this quantity by means of the HEOM.
Before we detail the HEOM, in this subsection, we first discuss expansions of $L(t)$.

We begin with the following expansion for a quantity $\exp(-i \omega t)$,
\begin{equation}
\label{eta expansion u}
e^{-i \omega t} = \sum_{k=1}^K \eta_k (\omega) u_k(t),
\end{equation}
with a set of functions $\{ u_k(t) \}_{k=1,2,\dots,K}$, which is closed under differentiation,
\begin{equation}
\label{u derivative}
\frac{d}{dt} u_k(t) = \sum_{k^{\prime} = 1}^{K} C_{k, k^{\prime}} u_{k^{\prime}}(t).
\end{equation}
In Eq.$\,$(\ref{eta expansion u}), $\eta_k(\omega)$ is the expansion coefficient.
This enables one to expand $L(t)$ Eq.$\,$(\ref{bath correlation compact}),
\begin{equation}
\label{bath correlation 2u}
\frac{1}{\hbar} L(t_1-t_2) = \sum_{k,k^{\prime}=1}^K D_{k, k^{\prime}} u_k(t_1) u_{k^{\prime}}^*(t_2),
\end{equation}
with
\begin{equation}
\label{D def}
D_{k, k^{\prime}} = \frac{1}{\hbar} \int_{-\infty}^{\infty} d\omega \frac{J(\omega)}{1-e^{-\beta \hbar \omega}} \eta_k (\omega) \eta_{k^{\prime}}^* (\omega). 
\end{equation}
Note that the matrix $D$ is hermitian and positive definite.

For later purposes (see Sec.$\,$\ref{method-higher}), we introduce a new basis which diagonalizes the matrix $D$. 
Since the matrix $D$ is hermitian, it can be diagonalized with a unitary matrix $U$,
\begin{equation}
\label{D diag}
D_{k, k^{\prime}} = \sum_{q=1}^{K} \lambda_q U_{k, q} U_{k^{\prime}, q}^{*},
\end{equation}
where $\{ \lambda_k \}$ are real and positive eigenvalues of $D$. 
With this basis, Eqs.$\,$(\ref{u derivative}) and (\ref{bath correlation 2u}) are transformed to
\begin{equation}
\label{v derivative}
\frac{d}{dt} v_k(t) = \sum_{k^{\prime} = 1}^{K} \bar{C}_{k, k^{\prime}} v_{k^{\prime}}(t),
\end{equation}
and
\begin{equation}
\label{bath correlation 2v}
\frac{1}{\hbar} L(t_1-t_2) = \sum_{k = 1}^{K} \lambda_k v_k(t_1) v_{k}^{*}(t_2),
\end{equation}
respectively, with $v_{k}(t) = \sum_{k^{\prime}=1}^{K} U_{k^{\prime}, k} u_{k^{\prime}}(t)$ and $\bar{C}_{k, k^{\prime}} = \sum_{q,q^{\prime} = 1}^{K} U_{q, k} C_{q, q^{\prime}} U_{q^{\prime}, k^{\prime}}^{*}$. By setting $t_1 = t$ and $t_2 = 0$ in Eq.$\,$(\ref{bath correlation 2v}), one further obtains
\begin{equation}
\label{bath correlation v}
\frac{1}{\hbar} L(t) = \sum_{k=1}^{K} \bar{c}_k v_k(t),
\end{equation}
with $\bar{c_k} = \lambda_k v_k^*(0)$.

An expansion of the form of Eq.$\,$(\ref{bath correlation v}) together with Eq.$\,$(\ref{v derivative}) was proposed in Ref.$\,$\cite{TOGHW15}, to extend applicability of the HEOM approach to problems where the initial bath is at zero temperature and to problems with various spectral densities.
In addition, our method further requires the relation Eq.$\,$(\ref{bath correlation 2v}).

\subsection{HEOM}
\label{method-HEOM}

In this subsection, we derive the HEOM, with which $\psi_0$ in Eq.$\,$(\ref{psi time evolution 0}) can be computed.
As in the conventional approach, we shall introduce a set of functions.
In the next subsection, we will show that they can be used to estimate the higher order contributions in the Taylor series of Eq.$\,$(\ref{g taylor first}).

With the expansion of $L(t)$ Eq.$\,$(\ref{bath correlation v}), $f[Q,t]$ in Eq.$\,$(\ref{f def}) is given by
\begin{equation}
\label{f with v}
f[Q,t] = \exp \left(- \sum_{k=1}^{K} \bar{c}_k \int_0^t ds \int_0^{s} d\tau \ h(Q(s)) h(Q(\tau)) v_{k}(s-\tau) \right).
\end{equation}
Thus, the time derivative of $\psi_0$ contains a term,
\begin{equation}
\label{psi time derivative unknown}
\begin{gathered}
- h(q_a) \int dq_c \ \varphi(q_c) \int_{(q_c,0)}^{(q_a,t)} D[Q] \, e^{ i S_S[Q,t] / \hbar } \\
\times f[Q,t] \ \sum_{k=1}^{K} \bar{c}_k \int_0^{t} ds \ h(Q(s)) v_k(t-s),
\end{gathered}
\end{equation}
whose time evolution is unknown. 
To deal with it, following the ideas of the HEOM approach, we introduce a set of functions,
\begin{equation}
\label{ef}
\begin{gathered}
\psi_{j_1,\dots,j_K}^{(n)} (q_a,t) = \int dq_c \ \varphi(q_c) \\
\times \int_{(q_c,0)}^{(q_a,t)} D[Q] \, e^{i S_S[Q,t] / \hbar } f[Q,t] \prod_{k=1}^{K} \frac{1}{i^{j_k}\sqrt{j_k !}} \{ y_k[Q,t] \}^{j_k},
\end{gathered}
\end{equation}
with $j_k = 0,1,2,\dots$ and $n = \sum_{k=1}^{K} j_k$. 
$y_k[Q,t]$ in this equation is defined as
\begin{equation}
\label{Y}
y_k[Q,t] = \int_0^t ds \ h(Q(s)) v_k(t-s).
\end{equation}
In the definition of $\psi_{j_1,\dots,j_K}^{(n)}$, we have multiplied a factor $\left(\prod_{k=1}^{K} i^{j_k} \sqrt{j_k !} \right)^{-1}$ for later discussions (see Sec.$\,$\ref{link}).

Using Eq.$\,$(\ref{v derivative}), one can derive the HEOM for $\psi_{j_1,\dots,j_K}^{(n)}$ (see \ref{derive HEOM}),
\begin{equation}
\label{HEOM}
\begin{gathered}
i \hbar \frac{\partial}{\partial t} \psi_{j_1,\dots,j_K}^{(n)} (q_a,t) = \\
\left\{ H_S(q_a) + i \hbar \sum_{k = 1}^{K} j_k \bar{C}_{k, k} \right\} \psi_{j_1,\dots,j_K}^{(n)} (q_a,t) \\
+ i \hbar \sum_{k \neq k^{\prime} = 1}^{K} \sqrt{j_k(j_{k^{\prime}}+1)} \ \bar{C}_{k, k^{\prime}} \ \psi_{j_1,\dots,j_k-1,\dots,j_{k^{\prime}}+1,\dots,j_K}^{(n)} (q_a,t) \\
+ h(q_a) \sum_{k=1}^{K} \sqrt{j_k} \ \hbar v_k(0) \ \psi_{j_1,\dots,j_k-1,\dots,j_K}^{(n-1)} (q_a,t) \\
+ h(q_a) \sum_{k=1}^{K} \sqrt{j_k+1} \ \hbar \bar{c}_k \ \psi_{j_1,\dots,j_k+1,\dots,j_K}^{(n+1)} (q_a,t). \\
\end{gathered}
\end{equation}
This equation is hierarchical with respect to $n$.
Since $y_k[Q,t=0] = 0$ (see Eq.$\,$(\ref{Y})), the initial conditions of $\psi_{j_1,\dots,j_K}^{(n)}$ read $\psi_{0,\dots,0}^{(0)} (q_a,t=0) = \varphi(q_a)$ and $\psi_{j_1,\dots,j_K}^{(n)} (q_a,t=0) = 0$ for $n \geqq 1$.
By solving the coupled equations of motion Eq.$\,$(\ref{HEOM}) with these initial conditions, one can derive the time evolution of $\psi_0$ as $\psi_0(q_a,t) = \psi_{0,\dots,0}^{(0)}(q_a,t)$. Then, one can calculate the lowest order contribution to the reduced density matrix with Eq.$\,$(\ref{rhoS time evolution 0}).

Such functions as Eq.$\,$(\ref{ef}) are called auxiliary functions in the conventional approach, because they are not directly used in calculating the reduced density matrix.
As we have shown,  they are auxiliary in terms of deriving $\psi_0$ and the 0-th order contribution of the Taylor series.
However, we will use the whole set of $\psi_{j_1,\dots,j_K}^{(n)}$ when estimating the reduced density matrix including the higher order contributions (see Eq.$\,$(\ref{rhoS time evolution phonon}) below).
Therefore, we should rather call them expansion functions.
We will give detailed discussions on the expansion functions in Sec.$\,$\ref{link}.

\subsection{Higher order contributions}
\label{method-higher}

In this subsection, we develop a scheme to estimate the higher order contributions in the Taylor expansion Eq.$\,$(\ref{g taylor first}).
Notice first that the exponent of Eq.$\,$(\ref{g def}) can be written in a form of 
\begin{equation}
\label{g phase before}
\frac{1}{\hbar} \int_0^t ds \int_0^t ds^{\prime} h(Q(s)) h(Q^{\prime}(s^{\prime})) L( (t-s) - (t-s^{\prime}) ).
\end{equation}
For $L/\hbar$ in this equation, we employ the expansion Eq.$\,$(\ref{bath correlation 2v}) and obtain
\begin{equation}
\label{g phase after}
\begin{gathered}
\sum_{k=1}^{K} \lambda_k \int_0^t ds \int_0^t ds^{\prime} h(Q(s)) h(Q^{\prime}(s^{\prime})) v_k(t-s) v_k^{*}(t-s^{\prime}) \\
= \sum_{k=1}^{K} \lambda_k y_{k}[Q,t] y_{k}^{*}[Q^{\prime},t].
\end{gathered}
\end{equation}
This leads to the Taylor expansion of $g[Q,Q^{\prime},t]$,
\begin{equation}
\label{g taylor v}
g[Q,Q^{\prime},t] = \sum_{n=0}^{\infty} \frac{1}{n!} \left\{ \sum_{k=1}^{K} \lambda_k y_{k}[Q,t] y_{k}^{*}[Q^{\prime},t] \right\}^n.
\end{equation}
It is useful to transform it into an occupation number representation,
\begin{equation}
\label{g taylor v occu}
g[Q,Q^{\prime},t] = \sum_{n=0}^{\infty} \ \sum_{(j_1+\dots+j_K = n)} \ \prod_{k = 1}^{K} \frac{1}{j_k!} \left\{ \lambda_k  y_{k}[Q,t] y_{k}^{*}[Q^{\prime},t] \right\}^{j_k},
\end{equation}
where $\sum_{(j_1+\dots+j_K = n)}$ means a sum over all possible configurations of $j_1,\dots,j_K$ with a constraint $j_1+\dots+j_K = n$. Then, the influence functional reads
\begin{equation}
\label{CL if taylor}
\begin{gathered}
\mathcal{F}[Q,Q^{\prime},t] = \sum_{n=0}^{\infty} \ \sum_{(j_1+\dots+j_K = n)} \left\{ \prod_{k = 1}^{K} \lambda_k^{j_k} \right\} \\
\times \left[ f[Q,t] \prod_{k = 1}^{K} \frac{1}{i^{j_k} \sqrt{j_k!}} \left\{ y_{k}[Q,t] \right\}^{j_k} \right] \left[ f[Q^{\prime},t] \prod_{k = 1}^{K} \frac{1}{i^{j_k} \sqrt{j_k!}} \left\{ y_{k}[Q^{\prime},t] \right\}^{j_k} \right]^{*}. \end{gathered}
\end{equation}
Finally, substituting this equation into Eq.$\,$(\ref{rhoS time evolution}) gives the time evolution of the reduced density matrix with the expansion functions defined by Eq.$\,$(\ref{ef}),
\begin{equation}
\label{rhoS time evolution phonon}
\begin{gathered}
\rho_S(q_a,q_b,t) =  \sum_{n=0}^{\infty} \ \sum_{(j_1+\dots+j_K = n)} \\ 
\times \left\{ \prod_{k = 1}^{K} \lambda_k^{j_k} \right\} \psi_{j_1,\dots,j_K}^{(n)}(q_a,t) \left\{ \psi_{j_1,\dots,j_K}^{(n)}(q_b,t) \right\}^*.
\end{gathered}
\end{equation}
In practical applications, one needs to truncate the sum at $n = N_{\rm max}$.
Then, this formula enables one to take into account up to $N_{\rm max}$-th order expansions of the Taylor series of $g[Q,Q^{\prime},t]$, by solving the HEOM Eq.$\,$(\ref{HEOM}) up to $n=N_{\rm max}$.
Such calculations include $\sum_{n=0}^{N_{\rm max}} (n+K-1)!/n!(K-1)!$ expansion functions. 
In the conventional approach, there exist several interpretations and corresponding schemes of truncating the hierarchy, including the delta function limit \cite{IT05} and the perturbation approximation \cite{SSK07,JWY18}.
In the next section, we will show that our formalism corresponds to taking into account up to $N_{\rm max}$-phonon states of the bath degrees of freedom.


\section{Interpretation of the method: phonon number representation}
\label{inter}

In the previous section, we have shown that the time evolution of the reduced density matrix is obtained with Eq.$\,$(\ref{rhoS time evolution phonon}), where $\psi_{j_1,\dots,j_K}^{(n)}$ satisfy the HEOM Eq.$\,$(\ref{HEOM}). 
In this section, we will give a physical interpretation of this method, that is, the expansion with respect to $n$ in Eq.$\,$(\ref{rhoS time evolution phonon}) is equivalent to the expansion of the harmonic oscillator bath with respect to the phonon number.

\subsection{Zero temperature}
\label{inter-zero}

Let us first consider a case where the temperature of the initial bath is absolute zero (that is, $\beta = \infty$).
For the sake of clarity, we begin with a single mode bath, 
\begin{equation}
\label{CL Hamiltonian single}
H_B + H_I = \hbar \omega_1 a_1^{\dagger} a_1 + h(q) d_1 (a_1 + a_1^{\dagger}).
\end{equation}

Introducing the eigenstates of $a_1^{\dagger} a_1$ as $\ket{n}_{n=0,1,\dots}$, the time evolution of the reduced density matrix can be written in the bracket form,
\begin{equation}
\label{rhoS time evolution single}
\begin{gathered}
\rho_S(q_a,q_b,t) = \sum_{n=0}^{\infty} \int dq_c \braket{ q_a, n | e^{-i H_{\rm tot} t / \hbar } | q_c, 0 } \varphi(q_c) \\
\times \left\{ \int dq_d \braket{ q_b, n | e^{-i H_{\rm tot} t / \hbar } | q_d, 0 } \varphi(q_d) \right\}^*,
\end{gathered}
\end{equation}
where we have used the fact that the initial bath is assumed to be at zero temperature, $\rho_B = \ket{0}\bra{0}$.

The definition of the reduced density matrix includes the trace operation over the bath degrees of freedom, ${\rm Tr}_B$, which corresponds to $\sum_{n=0}^{\infty}$ in Eq.$\,$(\ref{rhoS time evolution single}). 
Since we consider the initial condition at zero temperature, there is no phonon in the bath initially, while the number of phonon increases as the time goes on.
When the interaction is not so strong, or when one observes short time behaviors, the number of phonon in the bath would remain small. 
Therefore, it would be a reasonable approximation to truncate the sum in Eq.$\,$(\ref{rhoS time evolution single}) with a relatively small number, as is done in the coupled-channels method.

To implement the above expectation, we estimate Eq.$\,$(\ref{rhoS time evolution single}) at each $n$.
We find that the general form is given by
\begin{equation}
\label{n phonon expansion single}
\begin{gathered}
\braket{ q_a, n | e^{-i H_{\rm tot} t / \hbar } | q_c, 0 } = \\
\int_{(q_c,0)}^{(q_a,t)} D[Q] \, e^{i S_S[Q,t] / \hbar } f[Q,t] \frac{1}{i^n \sqrt{n!}} \left\{ z_1[Q,t] \right\}^n,
\end{gathered}
\end{equation}
with
\begin{equation}
\label{z def}
z_1[Q,t] = \int_0^{t} ds \ h(Q(s)) \frac{d_1}{\hbar} e^{-i \omega_1 (t-s)}.
\end{equation}
One sees that the exact form Eq.$\,$(\ref{rhoS time evolution}) with Eqs.$\,$(\ref{CL if}), (\ref{f def}), and (\ref{g def}) is reproduced when Eq.$\,$(\ref{n phonon expansion single}) is substituted into Eq.$\,$(\ref{rhoS time evolution single}) and the sum of $n$ is taken to $n = \infty$.
This leads one to an important conclusion in this paper; the phonon number expansion Eq.$\,$(\ref{rhoS time evolution single}) is equivalent to the Taylor expansion of $g[Q,Q^{\prime},t]$ (see Eq.$\,$(\ref{g taylor first})) at least when the bath has only a single mode.

This derivation can easily be extended to cases where the bath has several modes.
In such cases, Eq.$\,$(\ref{n phonon expansion single}) is modified to (see Eq.$\,$(\ref{prod if}))
\begin{equation}
\label{n phonon expansion several}
\begin{gathered}
\braket{ q_a, n_1, n_2, \dots | e^{-i H_{\rm tot} t / \hbar } | q_c, 0, 0, \dots } = \\
\int_{(q_c,0)}^{(q_a,t)} D[Q] \, e^{i S_S[Q,t] / \hbar} f[Q,t] \prod_i \frac{1}{i^{n_i} \sqrt{n_i !}} \left\{ z_i[Q,t] \right\}^{n_i},
\end{gathered}
\end{equation}
with $\ket{n_1,n_2,\dots} = \ket{n_1}\ket{n_2}\dots$, where $\ket{n_i}$ is the eigenstate of $a_i^{\dagger}a_i$.
$z_i$ is defined in a similar way to Eq.$\,$(\ref{z def}).
Hence, the $g[Q,Q^{\prime},t]$ part of the influence functional, Eq.$\,$(\ref{CL if}), reads
\begin{equation}
g[Q,Q^{\prime},t] = \sum_{n_1,n_2,\dots = 0}^{\infty}  \prod_i \frac{1}{n_i !} \left\{ z_i[Q,t] z_i^{*}[Q^{\prime},t] \right\}^{n_i}.
\end{equation}
This can be transformed into an expansion with respect to the total phonon number using the identity $\sum_{n_1,n_2,\dots = 0}^{\infty} = \sum_{n=0}^{\infty} \sum_{(n_1+n_2+\dots = n)}$,
\begin{equation}
\label{total phonon exp}
\begin{gathered}
\sum_{n=0}^{\infty} \sum_{(n_1+n_2+\dots = n)} \prod_i \frac{1}{n_i !} \left\{ z_i[Q,t] z_i^{*}[Q^{\prime},t] \right\}^{n_i} \\
= \sum_{n=0}^{\infty} \frac{1}{n!} \left\{ \sum_{i} z_i[Q,t] z_i^{*}[Q^{\prime},t] \right\}^{n}.
\end{gathered}
\end{equation}
Since $L(t)$ at zero temperature is given by $L(t) / \hbar = \sum_i d_i^2/\hbar^2 \ \exp(-i \omega_i t)$ (see Eqs.$\,$(\ref{bath correlation}) and (\ref{J})), one obtains
\begin{equation}
\sum_{i} z_i[Q,t] z_i^{*}[Q^{\prime},t] = \frac{1}{\hbar} \int_0^t ds \int_0^t ds^{\prime} h(Q(s)) h(Q^{\prime} (s^{\prime})) L(s^{\prime}-s).
\end{equation}
Therefore, the total phonon number expansion Eq.$\,$(\ref{total phonon exp}) is equivalent to the Taylor expansion Eq.$\,$(\ref{g taylor first}), even when the bath has more than one mode.

\subsection{Finite temperature}
\label{inter-finite}

The phonon number expansion Eq.$\,$(\ref{rhoS time evolution single}) is based on the initial bath at zero temperature, and obviously it cannot be applied to finite temperature problems.
At finite temperatures, the initial bath already contains several phonons, and thus the phonon number expansion may not be a reasonable approximation.

On the other hand, it is known that one can derive the time evolution of the reduced density matrix with the initial bath at finite temperature, by introducing additional degrees of freedom to the initial bath at zero temperature \cite{Wu18}.
To understand this based on the influence functional, let us rewrite $L(t)$ at finite temperatures, Eq.$\,$(\ref{bath correlation}), as
\begin{equation}
\label{bath correlation thermofield}
L(t) = \frac{1}{\hbar} \sum_i d_i^2 \left\{ n_{\beta}(\omega_i) + 1 \right\} e^{ -i \omega_i t} +  \frac{1}{\hbar} \sum_i d_i^2 n_{\beta}(\omega_i) e^{i \omega_i t},
\end{equation}
with the Bose-Einstein distribution function, $n_{\beta}(\omega) = (\exp(\beta \hbar \omega)-1)^{-1}$.
Comparing with $L(t)$ at zero temperature $L(t) = \sum_i d_i^2/\hbar \ \exp(-i \omega_i t)$, one can see that Eq.$\,$(\ref{bath correlation thermofield}) is reproduced with two kinds of independent bath at zero temperature.
The first term corresponds to a bath which couples to the system with the interaction strength $d_i \sqrt{n_{\beta}(\omega_i) + 1}$.
The second term, on the other hand, corresponds to a bath where the strength of interaction is $d_i \sqrt{n_{\beta}(\omega_i)}$.
Notice that this bath has a negative energy due to the sign of the phase of $\exp(i \omega_i t)$. 
From Eq.$\,$(\ref{prod if}), these baths are independent to each other.
Therefore, regarding the reduced density matrix, a finite temperature problem is equivalent to a zero temperature problem with the following Hamiltonian,
\begin{equation}
\label{thermo Hamiltonian}
\begin{gathered}
H_{\rm tot}^{(\beta)} = H_S(q,p) \\
+ \sum_{i} \hbar \omega_i \alpha_i^{\dagger} \alpha_i + h(q) \sum_i d_i \sqrt{n_{\beta}(\omega_i) + 1} \ (\alpha_i + \alpha_i^{\dagger}) \\
- \sum_{i} \hbar \omega_i \bar{\alpha}_i^{\dagger} \bar{\alpha}_i + h(q) \sum_i d_i \sqrt{n_{\beta}(\omega_i)} \ (\bar{\alpha}_i + \bar{\alpha}_i^{\dagger}),
\end{gathered}
\end{equation}
where $\alpha_i$, $\alpha_i^{\dagger}$ and $\bar{\alpha}_i$, $\bar{\alpha}_i^{\dagger}$ are the ladder operators for the first and the second baths, respectively.
In the limit of zero temperature ($\beta \to \infty$), one finds $n_{\beta}(\omega_i) \to 0$.
Hence, the Hamiltonian Eq.$\,$(\ref{thermo Hamiltonian}) is reduced to the original Hamiltonian Eq.$\,$(\ref{CL Hamiltonian}) by regarding $\alpha_i$ as $a_i$ and discarding the $\bar{\alpha}_i$ degrees of freedom which do not couple to the system in that limit.

In terms of phonon for the $\alpha_i$ and the $\bar{\alpha}_i$ degrees of freedom, the phonon number expansion is one approximate way to solve the reduced density matrix. 
Following exactly the same procedure as in the previous subsection, one reaches the same conclusion, that is, the Taylor expansion Eq.$\,$(\ref{g taylor first}) is equivalent to an expansion with respect to the number of phonon.
However, it should be kept in mind that these are quasi-phonons related to the $\alpha_i$ and the $\bar{\alpha}_i$ degrees of freedom.


\section{Practical application}
\label{num}

In this section, we apply the new method presented in the previous section to a concrete example.
To this end, we employ the Bessel functions for $\{ u_k \}$ to expand $\exp(-i \omega t)$ in Eq.$\,$(\ref{eta expansion u}).
Several advantages of this choice are discussed.
As a benchmark test of the method, we consider a quantum damped harmonic oscillator. We show that the exact results can be well reproduced with a parameter set with which effects of damping are observed.

\subsection{Choice of the physical dimension}
\label{num-dimension}

To begin with, let us clarify the physical dimension of quantities. 
We denote the dimension of energy and time as $[E]$ and $[T]$, respectively.
Firstly, we set $h(q)$ in Eq.$\,$(\ref{CL Hamiltonian}) be dimensionless.
This condition then leads to $d_i = [E]$, $J = [E]$, and $L/\hbar = [T^{-2}]$.

Secondly, we set $\{ u_k \}, \{ v_k \} = [T^{-1}]$ to make $\{ y_k[Q,t] \}$ be dimensionless quantities.
Then, this leads to $\{ \eta_k \} = [T]$ and $D$, $U$, and $\{ \lambda_k \}$ to be dimensionless.

\subsection{Use of the Bessel functions for $\{ u_k \}$}
\label{num-bessel}

In this subsection, we introduce the Bessel functions to expand $\exp(-i \omega t)$ in Eq.$\,$(\ref{eta expansion u}). 
For practical calculations, let us first introduce a cutoff parameter, $\Omega$, to the $\omega$-integral in the definition of $L(t)$, Eq.$\,$(\ref{bath correlation compact}), as
\begin{equation}
\label{bath correlation compact cutoff}
L(t) = \int_{-\Omega}^{\Omega} d\omega \frac{J(\omega)}{1-e^{-\beta \hbar \omega}} e^{-i \omega t}.
\end{equation}

Since the condition $|\omega/\Omega| \leqq 1$ holds in the domain of this integral, one can employ the Jacobi-Anger identity for the expansion of $\exp(-i \omega t)$ \cite{NT18,TC12,RK19}, that is,
\begin{equation}
\label{J-A identity}
e^{-i \omega t} = J_0(\Omega t) + 2 \sum_{n=1}^{\infty} (-i)^n T_{n}\left( \frac{\omega}{\Omega} \right) J_n(\Omega t),
\end{equation}
with the Chebyshev polynomials $T_n$ and the Bessel functions $J_n$. 
To follow the discussions in the previous subsection, we set $u_k(t) = \Omega J_{k-1}(\Omega t)$.
This leads to (see Eq.$\,$(\ref{eta expansion u}))
\begin{equation}
\label{eta for bessel}
\Omega \eta_k(\omega) = (2-\delta_{k,1}) (-i)^{k-1} T_{k-1} \left( \frac{\omega}{\Omega} \right),
\end{equation}
and
\begin{equation}
\label{bessel dt}
C_{k, k^{\prime}} = \left\{
\begin{array}{ll}
-\Omega & (k = 1, k^{\prime} = 2) \\
\Omega/2 & (k^{\prime} = k - 1) \\
-\Omega/2 & (k^{\prime} = k + 1) \\
0 & ({\rm else}).
\end{array}
\right.
\end{equation}

\begin{figure}[t]
\centering
\includegraphics[clip,width=6.5cm]{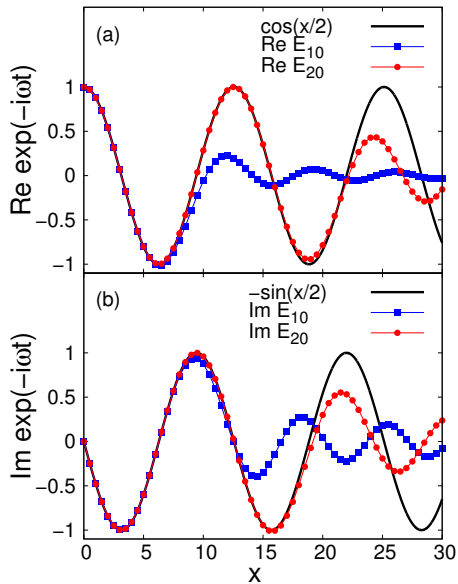}
\caption{
A comparison between the function $\exp(-i \omega t)$ and the right hand side of Eq.$\,$(\ref{J-A identity}) truncated at $n = K-1$,  $E_{K}(x) \equiv J_0(x) + 2\sum_{k=1}^{K-1} (-i)^k T_k(\omega/\Omega) J_k (x)$, for $\omega/\Omega = 1/2$ as a function of $x \equiv \Omega t$. 
The upper and the lower panels are for the real and the imaginary parts, respectively.
The solid lines denote the function $\exp(-i \omega t)$, while the solid lines with squares and the solid lines with circles are for $E_{10}$ and $E_{20}$, respectively. 
}
\label{fig:exp}
\end{figure}

There are certain advantages of using the Bessel functions for $\{ u_k \}$.
Firstly, since Eq.$\,$(\ref{J-A identity}) holds in general, one can apply the expansion to various kinds of spectral densities.
Another advantage, which was pointed out in Ref.$\,$\cite{NT18} and is rather important in practice, is related to a behavior of the Bessel functions as we discuss below.

In numerical calculations, it is impossible to take into account the infinite sum in Eq.$\,$(\ref{J-A identity}), and one needs a truncation with a cutoff (that is, $K$ in Eq.$\,$(\ref{eta expansion u})).
Notice that there are two requirements which should be satisfied with the truncation.
One is obviously that $\exp(-i \omega t)$ should be well reproduced.
To check this, let us denote the right hand side of Eq.$\,$(\ref{J-A identity}) with a truncation of the sum by $E_{K}(x) \equiv J_0(x) + 2\sum_{k=1}^{K-1} (-i)^k T_k(\omega/\Omega) J_k (x)$. 
In Fig.$\,$\ref{fig:exp}, we compare the real and the imaginary parts of $E_{10}(x)$ and $E_{20}(x)$ with those of $\exp(-i (\omega/\Omega) x)$, that is, $\cos(\omega x/\Omega)$ and $-\sin(\omega x/\Omega)$, respectively, for $\omega/\Omega = 1/2$.
It can be seen that the larger value of $K$ reproduces $\exp(-i \omega t)$ in the wider range, or for the longer time.

The other requirement is that a set $\{ u_k \}$ should be closed under differentiation (see Eq.$\,$(\ref{u derivative})).
However, regarding the Bessel functions, the chain of the differentiation Eq.$\,$(\ref{bessel dt}) continues infinitely.
Therefore, the truncation of the sum in Eq.$\,$(\ref{J-A identity}) might not be justified in this regard, even if $\exp(-i \omega t)$ is well reproduced.

In the original formalism of the HEOM, a sum of the exponential function of the form $\exp(-\gamma t)$ with a real quantity $\gamma$ was used to expand $L(t)$ (notice that this is different from an expansion of $\exp(-i \omega t)$ as is done in this paper)  \cite{TK89,YYLS04,IT05}.
A set of the exponential functions is obviously closed under differentiation.
One can extend this to orthogonal functions defined with positive arguments, $u_k(t) \equiv L_k(t) \exp(-t/2)$ with the Laguerre polynomials $\{ L_k(t) \}_{k = 0,1,\dots}$, because of the relation $du_k/dt (t) = -\sum_{n=0}^{k-1}u_n(t) - u_k(t)/2$ \cite{ZYS05}.

\begin{figure}[t]
\centering
\includegraphics[clip,width=8cm]{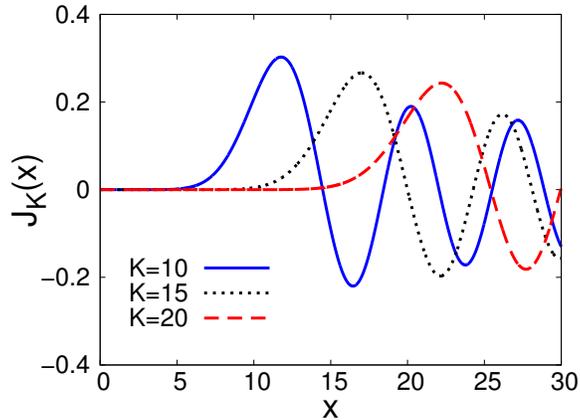}
\caption{The $K$-th order Bessel functions, $J_K(x)$.
The solid, the dotted, and the dashed lines are for $K = 10$, $15$, and $20$, respectively.}
\label{fig:bes}
\end{figure}

However, those expansions are not compatible with our method, and thus, it would be unavoidable to suffer from the aforementioned problem.
Fortunately, we can utilize the properties of the Bessel functions to deal with this problem. 
The leading order of the ascending series of the Bessel functions is given by $J_k(x) \sim x^k/2^k k!$.
From this, $x_{\epsilon}(k)$, which satisfies $J_k(x_{\epsilon}(k)) = \epsilon$ with a small positive number $\epsilon$, is given by $x_{\epsilon}(k) \sim 2 \sqrt[k]{\epsilon k!}$, which grows almost linearly with $k$.
Consequently, it is expected that the Bessel functions $\{ J_k (x) \}$ start growing at the larger $x$ for the larger $k$.
This is actually demonstrated in Fig.$\,$\ref{fig:bes}

Suppose that $J_{K}(\Omega s) \simeq 0$ holds for $0 \leqq s \leqq t$. 
Then, it is reasonable to approximate the time derivative of  $J_{K-1}(\Omega s)$ as $dJ_{K-1}(\Omega s)/ds \simeq \Omega J_{K-2} (\Omega s)/2$ in that range, and thus, the truncation of the expansion would work.
As $K$ is taken to be the larger, one can choose the larger $\Omega$ or one can calculate for the longer time, even though the numerical cost becomes more expensive, because the number of the expansion functions increases.
Note that $\Omega$ determines the characteristic time scale of the bath.
This implies that the method is suitable for problems where the time scale of the bath is comparable with that of the system, that is, for non-Markovian cases.

As was emphasized in Ref.$\,$\cite{NT18}, use of the Bessel functions is convenient especially when one takes the Ohmic spectral density with the circular cutoff,
\begin{equation}
\label{circular J}
J(\omega) = V_I \left( \frac{\omega}{\Omega} \right) \sqrt{1-\left(\frac{\omega}{\Omega}\right)^2},
\end{equation}
with $V_I$ being the strength of the interaction. 
This is because the imaginary part of $L(t)$ is given by ${\rm Im} L(t) = - \pi V_I \Omega / 8 \ (J_1 (\Omega t) + J_3(\Omega t))$ in this case, where ${\rm Im}$ denotes the imaginary part.

\subsection{Comparison of the numerical performance}
\label{num-comp}

Before going to a concrete application of our method, we would like to make several remarks on the numerical performance of the method in comparison with the existing HEOM methods.
It should be emphasized that each method has its own numerical advantages, and the most suitable method depends on problems to be discussed.
Here, we provide qualitative ideas which would be helpful to compare the numerical performance of each method.

With an expansion of $L(t)$ with the Bessel functions, one can utilize the extended version of the HEOM method introduced in Ref.$\,$\cite{TOGHW15}.
The method in Ref.$\,$\cite{TOGHW15} follows the time evolution of auxiliary density matrices, not vectors as in our method.
Thus, the numerical cost is quite sensitive to the dimension of the system, as has been discussed in Sec.$\,$\ref{intro}.
To compare the numerical cost comprehensively, however, one should also take into consideration the number of vectors or matrices required in computation.
As in our method, the method in Ref.$\,$\cite{TOGHW15} contains two cutoff numbers, that is, the maximum order of the hierarchy ($N_{\rm max}$ in our notation) and the number of the Bessel functions ($K$ in our notation).
With these parameters, the total number of matrices is given by $\sum_{n=0}^{N_{\rm max}} (n+K-1)!/n!(K-1)!$.
It is likely that our method requires a larger value of $N_{\rm max}$ to achieve convergence than the method in Ref.$\,$\cite{TOGHW15}, because we expand a part of the influence functional in the Taylor series (see Eq.$\,$(\ref{g taylor first})).
On the other hand, as was pointed out in Ref.$\,$\cite{NT18}, the method in Ref.$\,$\cite{TOGHW15} requires two times larger value of $K$ than our method, because one needs to construct the hierarchy with the forward path and the backward path simultaneously.
Note that $K$ is determined from the calculation time as discussed in the previous subsection, and $N_{\rm max}$ from the effective strength of the coupling (the strength of the coupling and the temperature of the initial bath, see Sec.$\,$\ref{num-dho}).
Therefore, our method would be more suitable for a large dimensional system and for a long time calculation, while the method in Ref.$\,$\cite{TOGHW15} would provide a better numerical performance for a strong coupling problem where a large number of phonon is excited.
Note that this does not necessarily mean that our method is perturbative calculation.
As has been discussed in Sec.$\,$\ref{inter}, our method is based on the coupled-channels approach with the phonon number representation of a bath.

One can also utilize the method of the hierarchical Schr\"odinger equations of motion introduced in Ref.$\,$\cite{NT18}.
It follows the time evolution of vectors as in our method.
The number of vectors involved in the method in Ref.$\,$\cite{NT18} is likely to be smaller than our method, since $N_{\rm max}$ can be smaller due to the same reasoning as discussed above, while $K$ is the same with the same setup. 
On the other hand, the method in Ref.$\,$\cite{NT18} requires twice longer calculation time than our method owing to the time integration for the forward path and for the backward path.
Therefore, once again, our method would be more suitable for a long time calculation, while the method in Ref.$\,$\cite{NT18} would be more convenient for a strong coupling problem.

\subsection{Damped harmonic oscillators}
\label{num-dho}

Let us now test our method using a quantum damped harmonic oscillator, in which the Hamiltonian for the system is given by
\begin{equation}
\label{HS dho}
H_S = \frac{p^2}{2M} + \frac{1}{2} M \omega_S^2 q^2 + h^2(q) \sum_i \frac{d_i^2}{\hbar \omega_i},
\end{equation}
where $M$ and $\omega_S$ are the mass and frequency of the system, respectively. 
The third term is the so called counter term.
With this term, the potential energy is understood as the one in the adiabatic limit, which has implicitly taken into account the couplings to the bath degrees of freedom \cite{CL83,THAB94}.
Introducing the oscillator length $q_S \equiv \sqrt{\hbar / M \omega_S}$, the interaction form factor $h(q)$ is taken to be $h(q) = q/q_S$.

Since the total Hamiltonian, Eq.$\,$(\ref{CL Hamiltonian 0}), is up to quadratic with respect to the coordinates and the momentums, the exact solution can be found by means of, for instance, the Laplace transform method \cite{BP02}.
Thus, this Hamiltonian provides an ideal opportunity for a benchmark calculation.
Several authors have applied their numerical methods to such systems, including the stochastic approach \cite{HL10} and the HEOM approach \cite{Tanimura15,XZZY15}. 
Especially, Ref.$\,$\cite{Tanimura15} is noteworthy to mention since the author of Ref.$\,$\cite{Tanimura15} extended the conventional approach to the thermalized initial condition with the total Hamiltonian where the system and the bath are correlated.
An extension to the imaginary time evolution was also successfully achieved, which enables one to calculate thermodynamic quantities.

Throughout numerical studies presented below, we use the spectral density given by Eq.$\,$(\ref{circular J}), and the expansion of $\exp(-i \omega t)$ with the Bessel functions, Eq.$\,$(\ref{J-A identity}).
We arbitrarily set $\hbar \omega_S = 2 \ {\rm eV}$, $V_I = 1 \ {\rm eV}$, and $\hbar \Omega = 4 \ {\rm eV}$. As will be shown, the damping of the amplitude can be seen with this parameter set. 
The initial wave function for the system is assumed to be of the Gaussian form,
\begin{equation}
\varphi(q) = \frac{1}{\sqrt[4]{2 \pi \sigma_0^2}} \, e^{-(q-q_0)^2 / 4 \sigma_0^2} \, e^{i p_0 q / \hbar},
\end{equation}
with $q_0/q_S = -1$, $\sigma_0/q_S = 1/\sqrt{2}$, and $p_0 q_S / \hbar = 0$. 
To solve the HEOM Eq.$\,$(\ref{HEOM}), we employ the fourth order Runge-Kutta method with the time grid $\Delta t / \hbar = 3.125 \times 10^{-3} \ {\rm eV}^{-1}$, and the space grid $\Delta q /q_S = 0.25$ in $-5.5 < q/q_S < 5.5$ (the dimension of the system reads $44$). 
We have confirmed that the results do not change significantly even if we use a smaller value of $\Delta q$ and/or a wider range of $q$.
In what follows, we show numerical results for both zero and finite temperature cases.

Let us first discuss the zero temperature case.
To this end, we first set $K = 10$. 
In other words, for the expansion of $\exp(-i \omega t)$, we include the Bessel functions up to $J_9$ in Eq.$\,$(\ref{J-A identity}).
The values of $\{ \lambda_k \}$ in Eq.$\,$(\ref{D diag}) are tabulated in Table$\,$\ref{tab:zero}.
Since $\lambda_1$ is so small, the expansion functions with nonzero $j_1$ have almost no contribution to the reduced density matrix (see Eq.$\,$(\ref{rhoS time evolution phonon})). 
Yet, we would keep them in the HEOM because of the $\bar{C}_{k, k^{\prime}}$ term in Eq.$\,$(\ref{HEOM}), that is, a set $\{ u_k \}$ should be closed under differentiation as discussed in Sec.$\,$\ref{num-bessel}.

\begin{table}[h]
\centering
\begin{tabular}{c|c}
\hline
\hline
$k$ & $\lambda_k$\\
\hline
$ \ \ \ \ \ 1 \ \ \ \ \ $ & $ \ \ \ \ \ 1.12 \times 10^{-14} \ \ \ \ \ $ \\
$ \ \ \ \ \ 2 \ \ \ \ \ $ & $ \ \ \ \ \ 2.18 \times 10^{-10} \ \ \ \ \ $ \\
$ \ \ \ \ \ 3 \ \ \ \ \ $ & $ \ \ \ \ \ 4.21 \times 10^{-7} \ \ \ \ \ $ \\
$ \ \ \ \ \ 4 \ \ \ \ \ $ & $ \ \ \ \ \ 1.84 \times 10^{-4} \ \ \ \ \ $ \\
$ \ \ \ \ \ 5 \ \ \ \ \ $ & $ \ \ \ \ \ 1.12 \times 10^{-2} \ \ \ \ \ $ \\
$ \ \ \ \ \ 6 \ \ \ \ \ $ & $ \ \ \ \ \ 3.13 \times 10^{-2} \ \ \ \ \ $ \\
$ \ \ \ \ \ 7 \ \ \ \ \ $ & $ \ \ \ \ \ 1.99 \times 10^{-1} \ \ \ \ \ $ \\
$ \ \ \ \ \ 8 \ \ \ \ \ $ & $ \ \ \ \ \ 2.61 \times 10^{-1} \ \ \ \ \ $ \\
$ \ \ \ \ \ 9 \ \ \ \ \ $ & $ \ \ \ \ \ 4.78 \times 10^{-1} \ \ \ \ \ $ \\
$ \ \ \ \ \ 10 \ \ \ \ \ $ & $ \ \ \ \ \ 5.16 \times 10^{-1} \ \ \ \ \ $ \\
\hline
\hline
\end{tabular}
\caption{The eigenvalues $\{ \lambda_k \}$ of the matrix, Eq.$\,$(\ref{D def}), at zero temperature.}
\label{tab:zero}
\end{table}

\begin{figure}[t]
\centering
\includegraphics[clip,width=11.58cm]{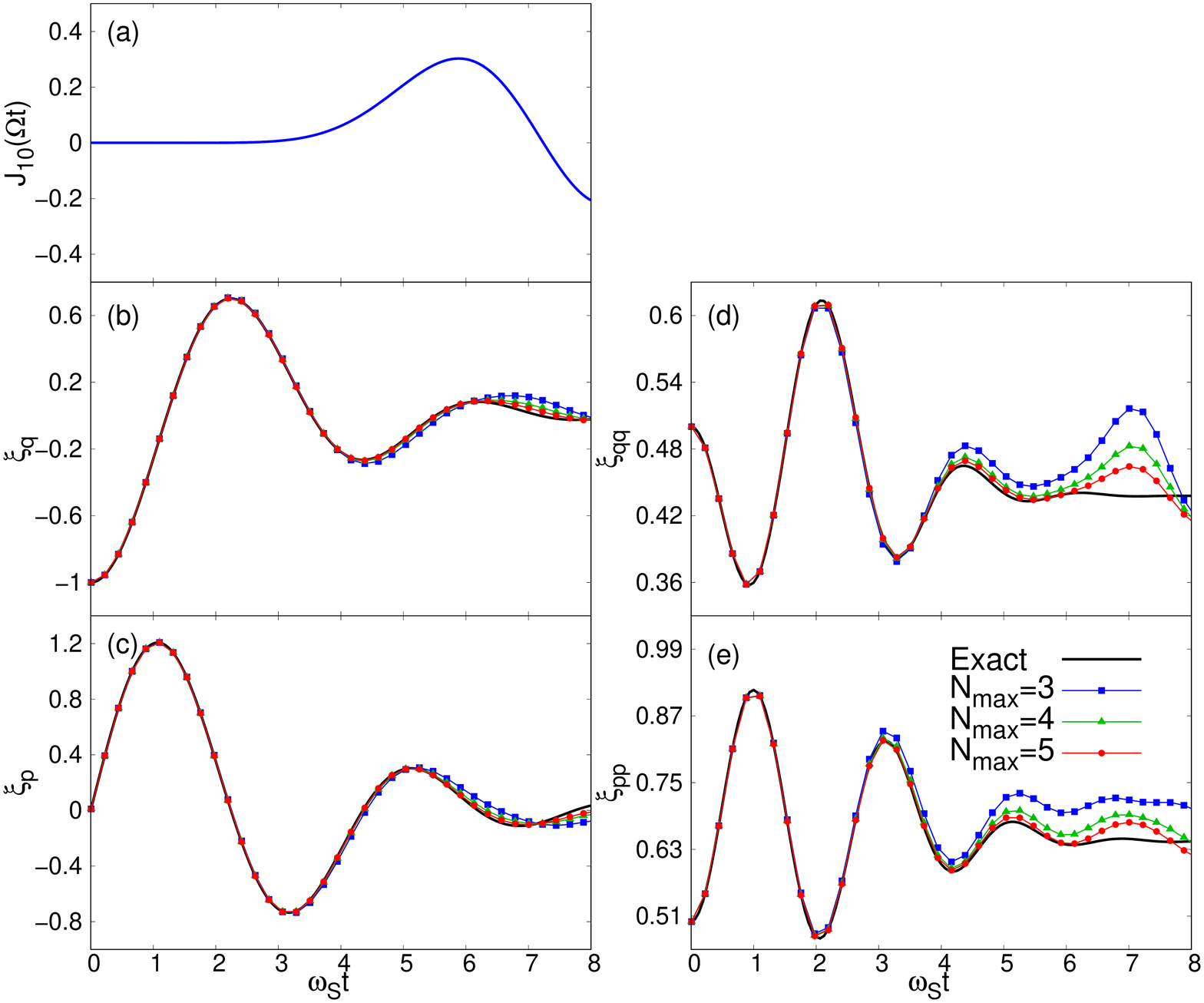}
\caption{Panel$\,$(a): The 10-th order Bessel function.
Panels$\,$(b)-(e): Comparison of the expectation values obtained with the Laplace transform method, which is supposed to be exact, to those with the HEOM with different $N_{\rm max}$.
The initial bath is assumed to be at zero temperature. The Bessel functions up to $J_9$ are included in the expansion in the HEOM method. 
$\xi_q$, $\xi_p$, $\xi_{qq}$, and $\xi_{pp}$ denote $\braket{q}/q_S$, $\braket{p} q_S /\hbar$, $\braket{\left( q-\braket{q} \right)^2}/q_S^2$, and $\braket{\left( p-\braket{p} \right)^2} q_S^2/\hbar^2$, respectively.
The solid lines show the exact results, while the solid lines with squares, triangles, and circles are the HEOM results with $N_{\rm max} = 3$, $4$, and $5$, respectively.}
\label{fig:-1pne}
\end{figure}

To test the applicability of our method, we compare expectation values obtained with the Laplace transform method, which is supposed to be exact, to those with the HEOM with several values of $N_{\rm max}$. 
We consider the following four expectation values: $\xi_q \equiv \braket{q}/q_S$, $\xi_p \equiv \braket{p} q_S /\hbar$, $\xi_{qq} \equiv \braket{\left( q-\braket{q} \right)^2}/q_S^2$, and $\xi_{pp} \equiv \braket{\left( p-\braket{p} \right)^2} q_S^2/\hbar^2$.
The results with $N_{\rm max} = 3$, $4$, and $5$ are shown in Fig.$\,$\ref{fig:-1pne}, which are compared to the exact results given by the solid lines.
We also show the behavior of $J_{10}(\Omega t)$ in Fig.$\,$\ref{fig:-1pne}(a), which is the least order among the neglected Bessel functions. 
According to the discussion in Sec.$\,$\ref{num-bessel}, the HEOM results are reliable up to $\omega_S t = 4 \sim 5$, at which $J_{10}(\Omega t)$ is negligibly small. 
One sees in Figs.$\,$\ref{fig:-1pne}(b) and (c) that $N_{\rm max} = 3$, $4$, and $5$ give similar results in this region, and they all agree well with the exact results. 
On the other hand, the results of $N_{\rm max} = 3$ and $N_{\rm max} = 4$ and $5$ somewhat deviate for $\omega_S t \geqq 3$ in Figs.$\,$\ref{fig:-1pne}(d) and (e), and only $N_{\rm max} = 4$ and $5$ reproduce the exact results.
This behavior is expected, because the larger number of phonon states are required to describe the more fine structures. 
The second order moments, $\xi_{qq}$ and $\xi_{pp}$, require more information of the reduced density matrix than the first order moments, $\xi_{q}$ and $\xi_{p}$, and thus, they need a larger model space to reproduce. 
As can be seen from this discussion, it should be kept in mind that a necessary value of $N_{\rm max}$ depends on physical quantities to be discussed.

For $N_{\rm max} = 3$, $4$, and $5$, the number of the expansion functions, Eq.$\,$(\ref{ef}), read 286, 1001, and 3003 with $K = 10$.
The calculations up to $\omega_S t = 5$ (800 time steps) with $25$ data points typically take 30 seconds, 2 minutes, and 4 minutes for $N_{\rm max} = 3$, $4$, and $5$ on a standard personal computer.
The first moments, $\xi_q$ and $\xi_p$, can be well reproduced with $N_{\rm max} = 3$.
Even when one is interested in the second moments, $\xi_{qq}$ and $\xi_{pp}$, $N_{\rm max} = 5$ leads to a good reproduction.

\begin{figure}[t]
\centering
\includegraphics[clip,width=6.5cm]{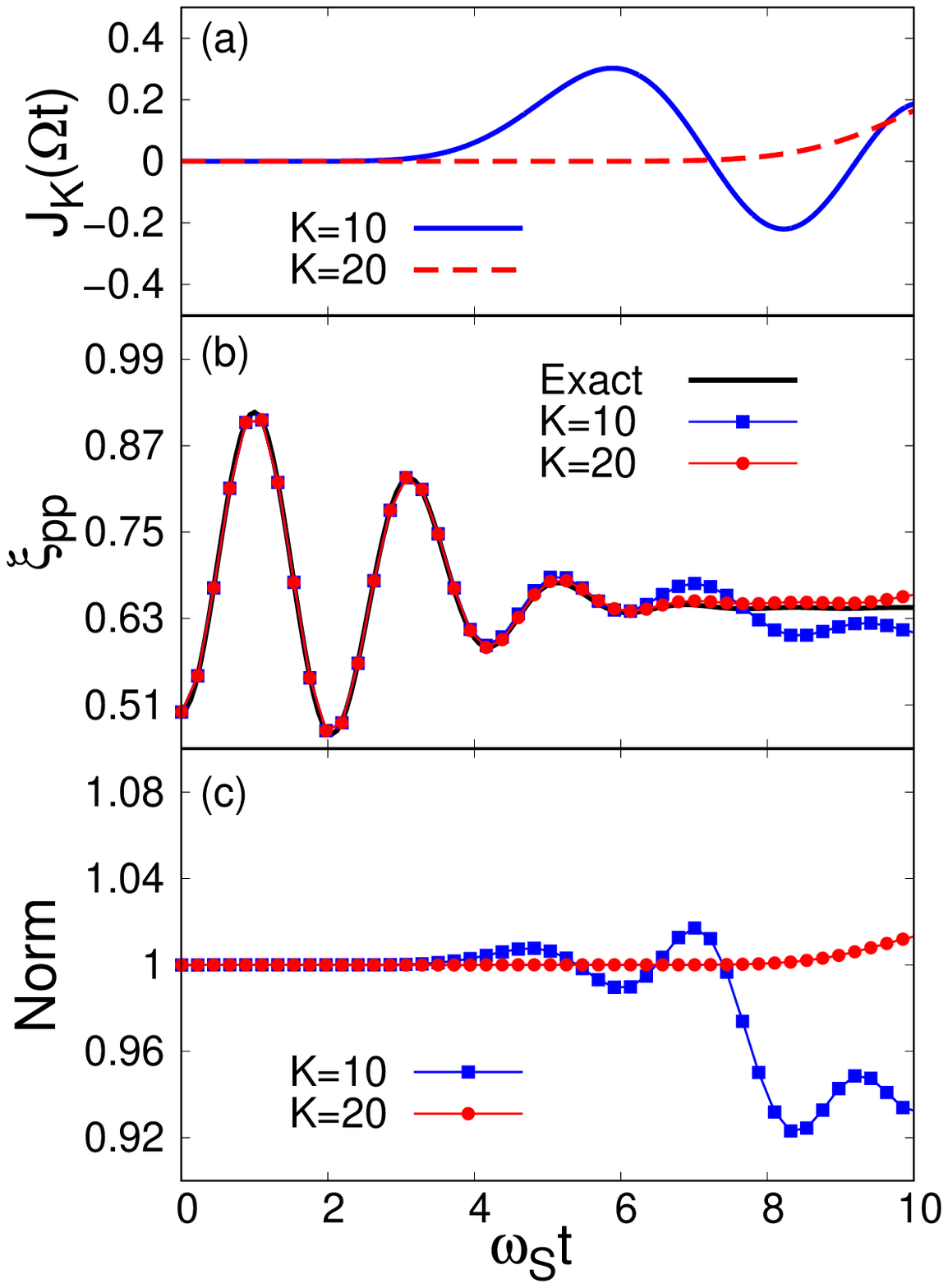}
\caption{Panel$\,$(a): The 10-th (the solid line) and the 20-th (the dashed line) order Bessel functions.
Panel$\,$(b): Comparison of the momentum width among the exact solution (the solid line), the HEOM with $K=10$ (the solid line with squares), and the HEOM with $K=20$ (the solid line with circles). For the HEOM calculations, $N_{\rm max} = 5$ is taken.
Panel$\,$(c): Comparison of the norm of the reduced density matrix. The solid line with squares is from the HEOM calculation with $K=10$, and the solid line with circles with $K=20$.
}
\label{fig:-1norm}
\end{figure}

\begin{figure}[t]
\centering
\includegraphics[clip,width=6.5cm]{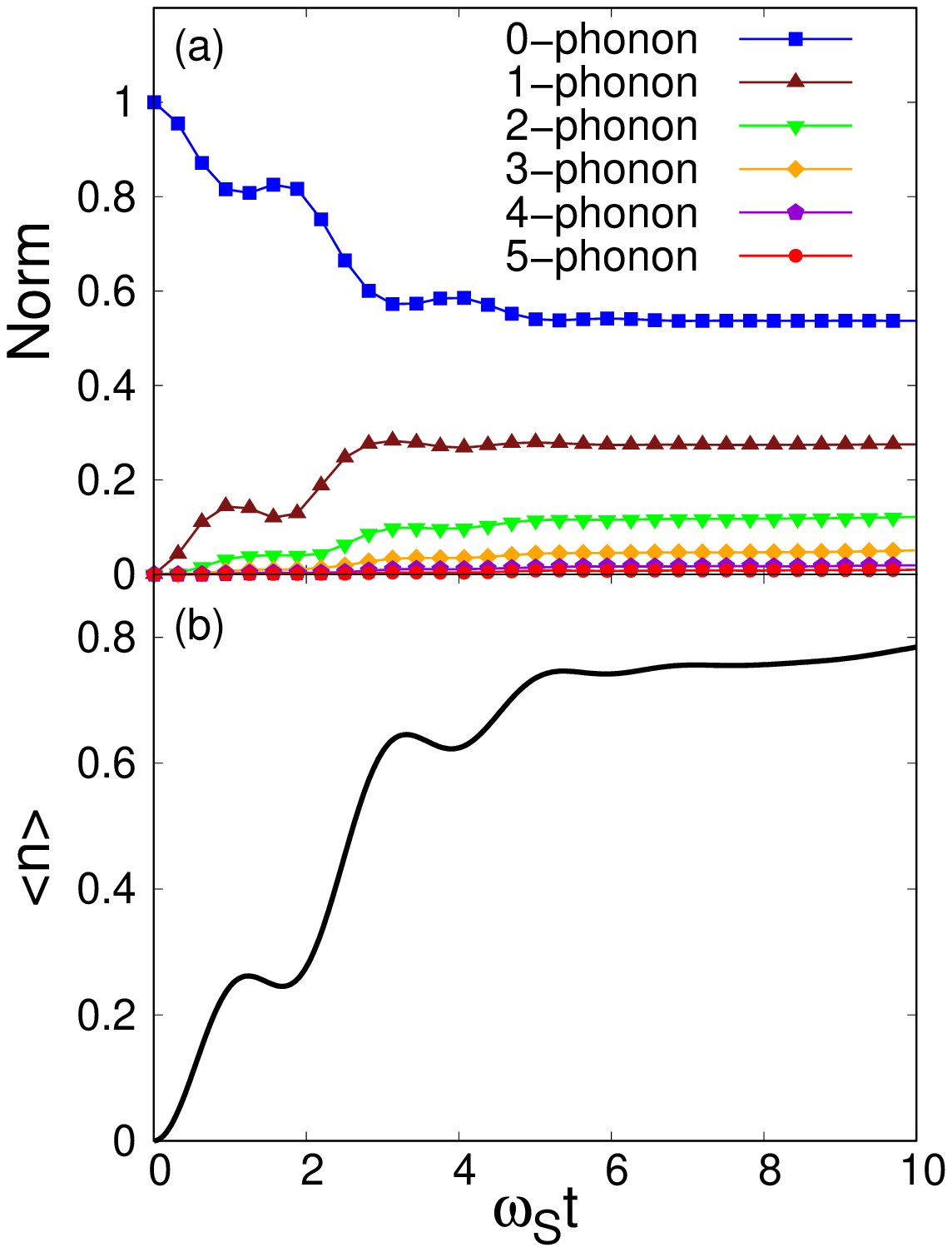}
\caption{Panel(a): Norm of the reduced density matrix for each phonon number, $n$. The solid line with squares, triangles, inverted triangles, diamonds, pentagons, and circles are for $n$ = 0, 1, 2, 3, 4, and 5, respectively. These are the results of the HEOM calculation with $K = 20$ and $N_{\rm max} = 5$.
Panel(b): Expectation value of the number of phonon.
}
\label{fig:-1phonon}
\end{figure}

For $\omega_S t \geqq 4 \sim 5$, the results of the HEOM with $K = 10$ deviate from the exact results.
It is likely that this originates from the fact that $J_{10}(\Omega t)$ is no longer negligible. 
One can cure this by taking a larger value of $K$, as discussed in Sec.$\,$\ref{num-bessel}. 
Fig.$\,$\ref{fig:-1norm}(b) compares the exact result for $\xi_{pp}$ to those with the HEOM with $K = 10$ and $20$. 
$N_{\rm max} = 5$ is chosen for the HEOM calculations.
The Bessel functions $J_{10}$ and $J_{20}$ are also plotted in Fig.$\,$\ref{fig:-1norm}(a).
One sees that the choice of $K=20$ can enlarge the applicability of the method. 
Actually, up to $\omega_S t \simeq 9$, it can closely follow the exact result of $\xi_{pp}$, which is the most difficult to describe among the expectation values under consideration.

In Fig.$\,$\ref{fig:-1norm}(c), we also plot the time dependence of the norm defined by ${\rm Tr}_S \rho_S$, with ${\rm Tr}_S$ being the trace operation over the degrees of freedom of the system.
Comparing with Fig.$\,$\ref{fig:-1norm}(a), one finds that the norm deviates from unity when the Bessel functions neglected in the expansion start having non-zero values. 
While the HEOM calculation works, the norm should be conserved. This point will become clearer in Sec.$\,$\ref{link}. 
Hence, the deviation of the norm from unity serves as a sign that the omitted Bessel functions start being non-negligible.

\begin{figure}[t]
\centering
\includegraphics[clip,width=11.58cm]{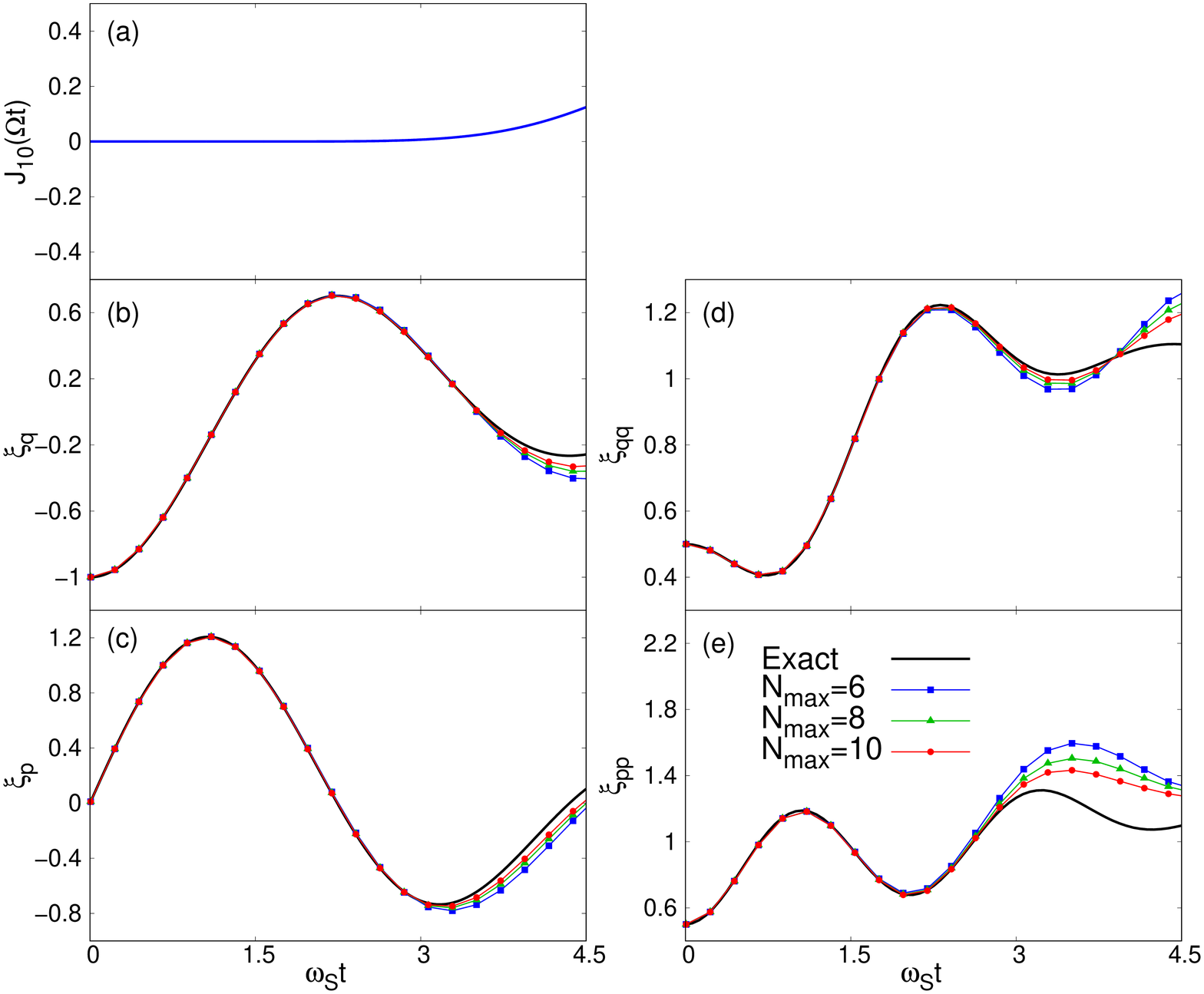}
\caption{Similar to Fig.$\,$\ref{fig:-1pne}, but at finite temperature with $\beta \hbar \Omega = 2$.}
\label{fig:2pne}
\end{figure}

\begin{figure}[t]
\centering
\includegraphics[clip,width=6.5cm]{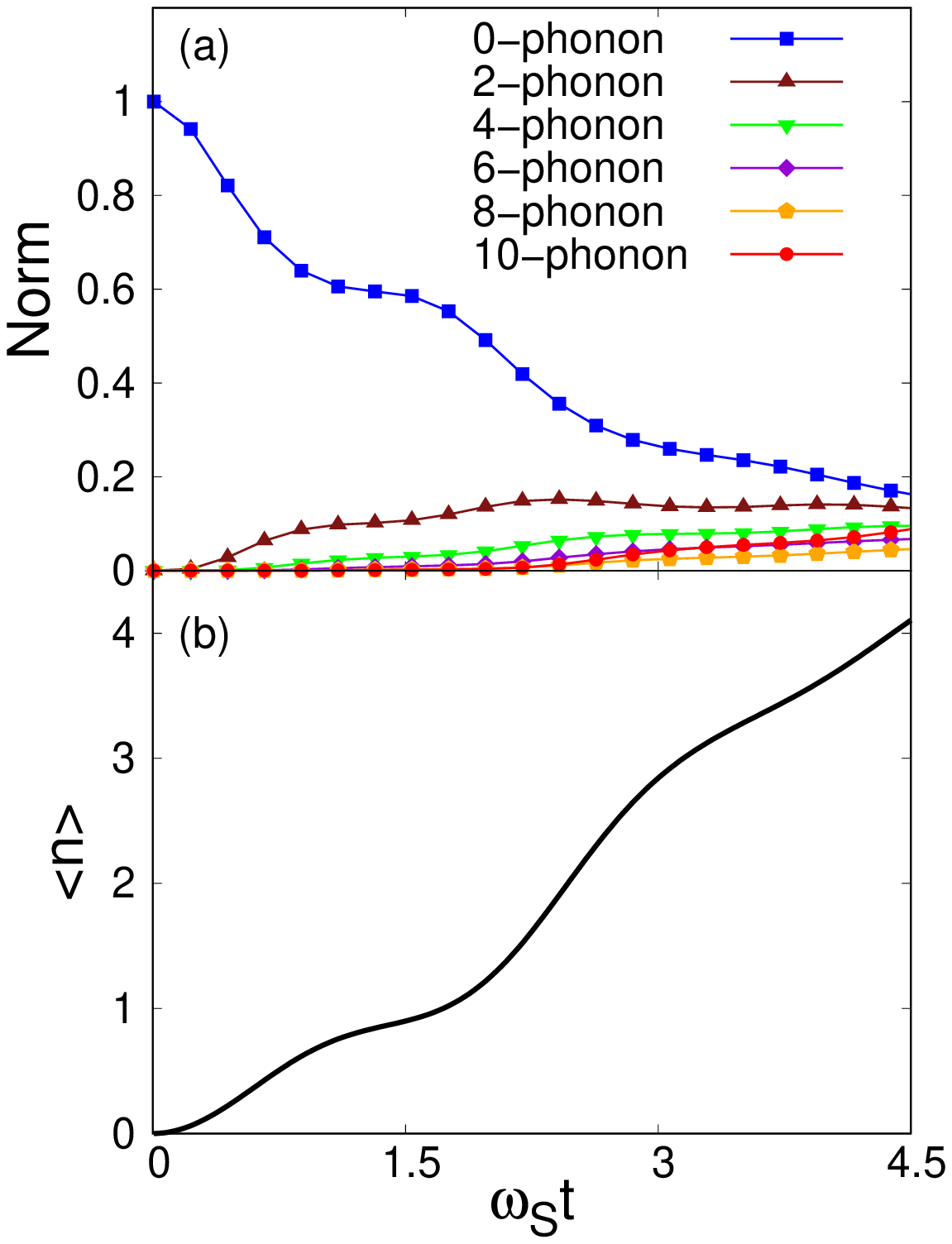}
\caption{Similar to Fig.$\,$\ref{fig:-1phonon}, but at finite temperature with $\beta \hbar \Omega = 2$.}
\label{fig:2phonon}
\end{figure}

In our method, in addition to the degrees of freedom of the system, we can also extract how much the bath is excited.
To see this, we calculate the norm of the reduced density matrix for each phonon number $n$. 
That is, writing the expansion of the reduced density matrix Eq.$\,$(\ref{rhoS time evolution phonon}) as $\rho_S \equiv \sum_{n=0}^{N_{\rm max}} \rho_S^{(n)}$, we compute it by ${\rm Tr}_S \rho_S^{(n)}$.
One can also calculate the expectation value of the phonon number as $\braket{n} \equiv \sum_{n=0}^{N_{\rm max}} n  \, {\rm Tr}_S \rho_S^{(n)} / {\rm Tr}_S \rho_S$. 
The result for $K=20$ and $N_{\rm max} = 5$ is shown in Fig.$\,$\ref{fig:-1phonon}.
As is expected, one can see in Fig.$\,$\ref{fig:-1phonon}(b) that the number of phonon in the bath gradually increases as the time goes on. 
On the other hand, as can be seen in Fig.$\,$\ref{fig:-1phonon}(a), the contribution of the $5$-phonon state is small in the whole time range.
This ensures that $N_{\rm max} = 5$ is sufficient to obtain reasonable results for the expectation values with the present parameter set. 
One can also see that the contribution of each phonon reaches its equilibrium at around $\omega_S t = 6$.
It is shown in Refs.$\,$\cite{Tanimura14,Tanimura15} that when the hierarchy elements stabilize they are equivalent to the thermal equilibrium state calculated with the imaginary-time HEOM. 
In our calculation, however, the expansion functions are still dependent on time even for $\omega_S t \geqq 6$.

Next, we apply our method to the initial bath at finite temperature with $\beta \hbar \Omega = 2$. 
As discussed in Sec.$\,$\ref{inter-finite}, phonon involved here is not a physical one.
Yet, we call it phonon throughout this study. 
As in the zero temperature case, we set $K = 10$. 
The values of $\{ \lambda_k \}$ are tabulated in Table$\,$\ref{tab:finite}. 
They have larger values compared to the zero temperature case shown in Table$\,$\ref{tab:zero}. 
Since the strength of interaction is described by $\{ \lambda_k \}$, this indicates that the temperature effectively strengthens the coupling.
This can also be seen from the definition of the thermo-Hamiltonian, Eq.$\,$(\ref{thermo Hamiltonian}).

\begin{table}[h]
\centering
\begin{tabular}{c|c}
\hline
\hline
$k$ & $\lambda_k$\\
\hline
$ \ \ \ \ \ 1 \ \ \ \ \ $ & $ \ \ \ \ \ 5.65 \times 10^{-3} \ \ \ \ \ $ \\
$ \ \ \ \ \ 2 \ \ \ \ \ $ & $ \ \ \ \ \ 3.24 \times 10^{-2} \ \ \ \ \ $ \\
$ \ \ \ \ \ 3 \ \ \ \ \ $ & $ \ \ \ \ \ 5.76 \times 10^{-2} \ \ \ \ \ $ \\
$ \ \ \ \ \ 4 \ \ \ \ \ $ & $ \ \ \ \ \ 1.57 \times 10^{-1} \ \ \ \ \ $ \\
$ \ \ \ \ \ 5 \ \ \ \ \ $ & $ \ \ \ \ \ 2.88 \times 10^{-1} \ \ \ \ \ $ \\
$ \ \ \ \ \ 6 \ \ \ \ \ $ & $ \ \ \ \ \ 3.43 \times 10^{-1} \ \ \ \ \ $ \\
$ \ \ \ \ \ 7 \ \ \ \ \ $ & $ \ \ \ \ \ 5.92 \times 10^{-1} \ \ \ \ \ $ \\
$ \ \ \ \ \ 8 \ \ \ \ \ $ & $ \ \ \ \ \ 6.25 \times 10^{-1} \ \ \ \ \ $ \\
$ \ \ \ \ \ 9 \ \ \ \ \ $ & $ \ \ \ \ \ 8.47 \times 10^{-1} \ \ \ \ \ $ \\
$ \ \ \ \ \ 10 \ \ \ \ \ $ & $ \ \ \ \ \ 8.72 \times 10^{-1} \ \ \ \ \ $ \\
\hline
\hline
\end{tabular}
\caption{The eigenvalues $\{ \lambda_k \}$ of the matrix, Eq.$\,$(\ref{D def}), at temperature of $\beta \hbar \Omega = 2$.}
\label{tab:finite}
\end{table}

Since the interaction becomes effectively stronger, one needs to take into account a larger number of phonon compared to the zero temperature case.
The results of the HEOM calculations with $K = 10$ ($i.e.$ an expansion up to $J_9$) and $N_{\rm max} = 6$, 8, and 10 are shown in Fig.$\,$\ref{fig:2pne}.
One finds that the first moments, $\xi_q$ and $\xi_p$, at $N_{\rm max}=10$ agree with the exact results in the range where $J_{10}(\omega_S t)$ is negligible. 
However, the results of the HEOM calculations for the second moments, $\xi_{qq}$ and $\xi_{pp}$, deviate from the exact results even in that region. 
This corresponds to that the number of phonon included in the calculation, $N_{\rm max} = 10$, is insufficient to describe the second moments with the strength of the interaction given in Table$\,$\ref{tab:finite}.
To see it more transparently, we plot the norm for each phonon number with $N_{\rm max} = 10$ in Fig.$\,$\ref{fig:2phonon}.
As can be seen, the contribution of the 10-phonon state is not negligible. 
This implies that a larger number of phonon states should be taken into account. 
Although the numerical calculation becomes more expensive, it should be emphasized that this can be systematically improved with the present method.


\section{Link to the total wave function}
\label{link}

We have introduced the expansion functions Eq.$\,$(\ref{ef}), which enable one to calculate the reduced density matrix.
In this section, inspired by the coupled-channels method, we provide their link to the total wave function. 
This consideration will lead us to an introduction of ladder operators which satisfy the boson commutation relation.
In terms of them, we will gain a clear idea on the reason why one can reduce the number of the relevant degrees of freedom of the bath with the expansion Eq.$\,$(\ref{eta expansion u}).
Throughout this section, we consider the initial bath at zero temperature.

\subsection{Discrete bath}
\label{link-discrete}

For a discrete bath, since $L(t)$ is given by (see Eqs.$\,$(\ref{bath correlation}) and (\ref{J}))
\begin{equation}
\frac{1}{\hbar}L(t_1-t_2) = \sum_i \left( \frac{d_i}{\hbar} e^{-i \omega_i t_1} \right) \left( \frac{d_i}{\hbar} e^{-i \omega_i t_2} \right)^*,
\end{equation}
one can take $v_i(t) = d_i/\hbar \ e^{-i \omega_i t}$ with $\lambda_i = 1$ in Eq.$\,$(\ref{bath correlation 2v}).
Let us denote the expansion functions, Eq.$\,$(\ref{ef}), in this case as $\phi_{n_1,n_2,\dots}^{(n)}$.
In the definition of the expansion functions, $\{ y_k[Q,t] \}$ is given by Eq.$\,$(\ref{Y}).
When one substitutes $v_i(t) = d_i/\hbar \ e^{-i \omega_i t}$ into Eq.$\,$(\ref{Y}), one obtains $\{ z_i[Q,t] \}$ defined by Eq.$\,$(\ref{z def}).
Therefore, $\phi_{n_1,n_2,\dots}^{(n)}$ can be written with $\{ z_i[Q,t] \}$ as
\begin{equation}
\label{ef discrete}
\begin{gathered}
\phi_{n_1,n_2,\dots}^{(n)} (q_a,t) = \int dq_c \ \varphi(q_c) \\
\times \int_{(q_c,0)}^{(q_a,t)} D[Q] \, e^{i S_S[Q,t] / \hbar } f[Q,t] \prod_{i} \frac{1}{i^{n_i}\sqrt{n_i !}} \{ z_i[Q,t] \}^{n_i},
\end{gathered}
\end{equation}
where $n_1,n_2,\dots$ satisfy $n_1 + n_2 + \dots = n$.
Comparing this to Eq.$\,$(\ref{n phonon expansion several}), one finds
\begin{equation}
\label{exp and total wf}
\begin{gathered}
\phi_{n_1,n_2,\dots}^{(n)} (q_a,t) = \int dq_c \braket{ q_a, n_1, n_2, \dots | e^{- i H_{\rm tot} t / \hbar } | q_c, 0, 0, \dots } \varphi(q_c) \\
\equiv \braket{ q_a, n_1, n_2, \dots| \Psi (t) },
\end{gathered}
\end{equation}
where $\ket{\Psi (t)}$ is the total wave function at time $t$, that is, 
\begin{equation}
\label{te Psi}
\ket{\Psi (t)} = e^{-i H_{\rm tot} t / \hbar} \ket{\Psi (t=0)}.
\end{equation}

This indicates that the expansion functions are nothing but the coefficients in an expansion of the total wave function with respect to the phonon eigenstates, 
\begin{equation}
\label{exp and total wf}
\braket{q_a|\Psi (t)} = \sum_{n=0}^{\infty} \sum_{(n_1 + n_2 + \dots = n)} \phi_{n_1,n_2,\dots}^{(n)} (q_a,t) \ \ket{n_1, n_2, \dots}.
\end{equation}
Actually, one can derive the HEOM, Eq.$\,$(\ref{HEOM}), by substituting this definition into the Schr\"odinger equation. 
The relation to the reduced density matrix, Eq.$\,$(\ref{rhoS time evolution phonon}), is also obtained from this, since $\lambda_i = 1$ for all $i$.

\subsection{Ladder operators for $\psi_{j_1,\dots,j_K}^{(n)}$}
\label{link-continuous}

In the previous subsection, we have set $v_i(t) = d_i/\hbar \ e^{-i \omega_i t}$.
Obviously, this is not the only choice.
In Sec.$\,$\ref{num-bessel}, for instance, we have discussed the advantages of employing the Bessel functions for $\{ u_k \}$ in Eq.$\,$(\ref{eta expansion u}), and thus, for $\{ v_k \}$.
There may be other useful functions in this context. 
Let us denote the expansion functions defined with those functions as $\psi_{j_1,\dots,j_K}^{(n)}$.
It should be noticed that expanding $\exp(-i \omega t)$ with those functions enables one to establish a method whose numerical cost is independent of the number of the harmonic oscillator modes.
For instance, one can even deal with situations in which the spectral density $J(\omega)$ defined in Eq.$\,$(\ref{J}) is a continuous function of $\omega$, as has been done in Sec.$\,$\ref{num-dho}.

In the previous subsection, we have shown that $\phi_{n_1,n_2,\dots}^{(n)}$ can be interpreted as the coefficients in an expansion of the total wave function.
One can attach a similar interpretation to $\psi_{j_1,\dots,j_K}^{(n)}$.
To see it, we first rewrite the expansion of $\exp(-i \omega t)$, Eq.$\,$(\ref{eta expansion u}), in terms of $v_k(t)$,
\begin{equation}
\label{eta expansion v}
e^{-i \omega_i t} = \sum_{k=1}^{K} \bar{\eta}_k (\omega_i) v_k (t),
\end{equation}
with $\bar{\eta}_k (\omega_i) = \sum_{k^{\prime} = 1}^K U_{k^{\prime}, k}^* \eta_{k^{\prime}} (\omega_i)$. 
From the definition of $\lambda_k$, Eq.$\,$(\ref{D diag}), one finds
\begin{equation}
\label{eta vs lambda}
\sum_i \frac{d_i^2}{\hbar^2} \bar{\eta}_k(\omega_i) \bar{\eta}_{k^{\prime}}^* (\omega_i) = \delta_{k, k^{\prime}} \lambda_k.
\end{equation}

Substituting Eq.$\,$(\ref{eta expansion v}) into the definition of $z_i[Q,t]$ gives the relation between $z_i[Q,t]$ and $y_k[Q,t]$,
\begin{equation}
\label{z vs y}
z_i[Q,t] = \frac{d_i}{\hbar} \sum_{k=1}^{K} \bar{\eta}_k (\omega_i) y_k[Q,t],
\end{equation}
which leads to the relation between $\phi_i \equiv \phi_{n_i = 1, n_j = 0 (j \neq i)}^{(1)}$ and $\psi_k \equiv \psi_{j_k = 1, j_q = 0 (q \neq k)}^{(1)}$,
\begin{equation}
\label{phi vs psi}
\phi_i(q_a,t) = \frac{d_i}{\hbar} \sum_{k=1}^{K} \bar{\eta}_k (\omega_i) \psi_k(q_a,t).
\end{equation}
Hence, the single phonon state in Eq.$\,$(\ref{exp and total wf}) can be transformed to an expansion with respect to $\psi_k$,
\begin{equation}
\label{phi vs psi single}
\begin{gathered}
\sum_i \phi_i (q_a,t) a_i^{\dagger} \ket{0} =  \sum_{k=1}^{K} \psi_k(q_a,t) \left( \sum_i \frac{d_i}{\hbar} \bar{\eta}_k(\omega_i) a_i^{\dagger} \right) \ket{0} \\
\equiv \sum_{k=1}^{K} \psi_k(q_a,t) b_k^{\dagger} \ket{0}.
\end{gathered}
\end{equation}
with the vacuum $\ket{0} = \ket{0,0,\dots}$.
Here, we have introduced operators $b_k^{\dagger} \equiv \sum_i d_i \bar{\eta}_k(\omega_i) a_i^{\dagger} / \hbar$ for $k = 1,\dots,K$, which, from the commutation relation of $\{ a_i \}$ and $\{ a_i^{\dagger} \}$ together with Eq.$\,$(\ref{eta vs lambda}), satisfy the following commutation relations,
\begin{equation}
\label{b commutation}
\begin{gathered}
\left[b_{k}, b_{k^{\prime}} \right] = \left[ b_{k}^{\dagger}, b_{k^{\prime}}^{\dagger} \right] = 0, \\
\left[ b_{k}, b_{k^{\prime}}^{\dagger} \right] = \delta_{k, k^{\prime}} \lambda_k.
\end{gathered}
\end{equation}
This implies that the $\{ b_k^{\dagger} \}$ operators describe a creation of boson.
In addition, one finds
\begin{equation}
\label{b vs a sum}
\hbar \sum_{k=1}^K b_k^{\dagger} v_k (t) = \sum_i d_i a_i^{\dagger} e^{-i \omega_i t},
\end{equation}
which indicates that $\sum_{k = 1}^K b_k^{\dagger} (t) v_k (t)$ is time independent, where $b_k^{\dagger} (t) = \exp(i H_B t / \hbar) b_k^{\dagger} \exp(-i H_B t / \hbar)$ is the interaction picture of $b_k^{\dagger}$.

\begin{figure}[t]
\centering
\includegraphics[clip,width=6.5cm]{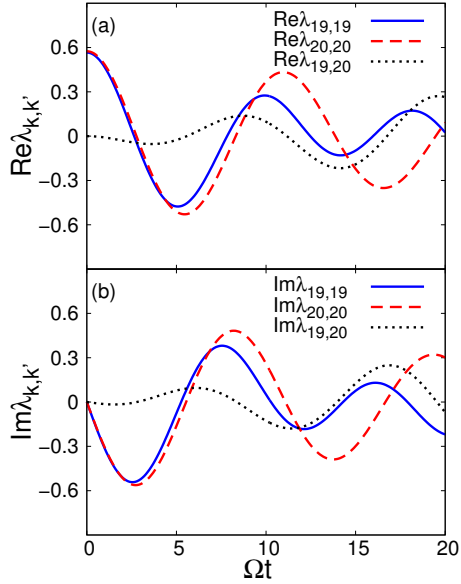}
\caption{
Panel(a): The real part of the commutation relations $\lambda_{k, k^{\prime}}(t) \equiv [ b_{k}(t), b_{k^{\prime}}^{\dagger}(0) ]$ with the expansion of $\exp(-i \omega t)$ with the Bessel functions up to $K = 20$ (see Eqs.$\,$(\ref{eta expansion u}) and (\ref{J-A identity})). The label $k$ is in the increasing order of the values of $\{ \lambda_k \}$, and we plot the two largest values of $k$, that is, $\lambda_{19,19}(t)$ (the solid line), $\lambda_{20,20}(t)$ (the dashed line), and $\lambda_{19,20}(t)$ (the dotted line).
Panel(b): Similar to Panel(a), but their imaginary parts are plotted.
}
\label{fig:lambda}
\end{figure}

To enlarge physical intuition of the quanta associated with the $\{ b_k \}$ and $\{ b_k^{\dagger} \}$ operators, let us examine the commutation relation between $b_{k}$ and $b_{k^{\prime}}^{\dagger}$ at different times,
\begin{equation}
\begin{gathered}
\lambda_{k, k^{\prime}}(t_1-t_2) \equiv \left[ b_{k}(t_1), b_{k^{\prime}}^{\dagger}(t_2) \right] \\
= \sum_i \frac{d_i^2}{\hbar^2} \bar{\eta}_{k^{\prime}} (\omega_i) \bar{\eta}_{k}^* (\omega_i) e^{-i \omega_i (t_1-t_2)} \\
= \frac{1}{\hbar} \int_0^{\infty} d\omega \ J(\omega) \bar{\eta}_{k^{\prime}} (\omega) \bar{\eta}_{k}^* (\omega) e^{-i \omega (t_1-t_2)},
\end{gathered}
\end{equation}
which is diagonal when the argument is 0, that is, $\lambda_{k, k^{\prime}} (0) = \delta_{k, k^{\prime}} \lambda_k$. It is related to $L(t)$ as $L(t)/\hbar = \sum_{k,k^{\prime} = 1}^K \lambda_{k, k^{\prime}}(t) v_{k^{\prime}}(0) v_k^*(0)$.

As an example, in Fig.$\,$\ref{fig:lambda}, we show $\lambda_{k,k^{\prime}}(t)$ with three different combinations of $k$ and $k^{\prime}$ with the same setup as in Sec.$\,$\ref{num-dho}.
That is, we take the Ohmic spectral density with the circular cutoff given by Eq.$\,$(\ref{circular J}) with $\hbar \Omega = 4$ eV and $V_I = 1$ eV.
To expand $\exp(-i \omega t)$ in Eq.$\,$(\ref{eta expansion u}), we use the Bessel functions, Eq.$\,$(\ref{J-A identity}), up to $K = 20$. 
The label $k$ is sorted out in the increasing order of the values of $\{ \lambda_k \}$ as in Tables$\,$\ref{tab:zero} and \ref{tab:finite}, and we focus on the two largest values of $k$, that is, $\lambda_{19,19}(t)$, $\lambda_{20,20}(t)$, and $\lambda_{19,20}(t)$.

In $\lambda_{19,19}(t)$ and $\lambda_{20,20}(t)$ in Fig.$\,$\ref{fig:lambda}, it can be seen that their oscillation patters are structured in such a way that the amplitudes vary as the time goes on.
Notice that this is in contrast to the phonons described by the $\{ a_i \}$ and $\{ a_i^{\dagger} \}$ operators, since the commutation relations oscillate with a constant amplitude, $[a_i(t_1),a_j^{\dagger}(t_2)] = \delta_{i,j} \exp(- i \omega_i (t_1 - t_2))$.

One can also see in Fig.$\,$\ref{fig:lambda} that the non-diagonal element, $\lambda_{19,20}(t)$, starts having non-zero values at a finite value of $t$. 
Notice that the non-diagonal elements appear when one considers the transition amplitude,
\begin{equation}
\braket{0|b_k \, e^{-i H_B t / \hbar} \, b_{k^{\prime}}^{\dagger} |0} = \lambda_{k, k^{\prime}}(t),
\end{equation}
which describes the amplitude of the $b_k$-mode at time $t$ with the initial condition $b_{k^{\prime}}^{\dagger}\ket{0}$.
Hence, the increase of the non-diagonal elements of $\lambda_{k, k^{\prime}}(t)$ indicates that the quanta associated with the $\{ b_k \}$ and $\{ b_k^{\dagger} \}$ operators can transform from one mode to another.
In other words, different modes interact with each other.
Therefore, the Caldeira-Leggett model can be viewed in such a way that a system couples to a bath, which is composed of finite modes of interacting bosons, even when the number of the harmonic oscillator modes is infinite.

A formulation of the HEOM based on quasiparticle was proposed in Ref.$\,$\cite{Yan14}, and the author of Ref.$\,$\cite{Yan14} call it {\it dissipaton}.
To see a connection with this, we denote the interaction Hamiltonian as $H_I = h(q) X$ with $X \equiv \sum_{i} d_i (a_i + a_i^{\dagger}) $. Eq.$\,$(\ref{b vs a sum}) enables one to decompose $X$ into each $\{ b_k \}$-mode as
\begin{equation}
\label{X decomposition}
X = \hbar \sum_{k=1}^{K} \left( v_k^*(0) b_k + v_k(0) b_k^{\dagger} \right).
\end{equation}
This can be regarded as a concrete form of the {\it dissipaton} decomposition in Ref.$\,$\cite{Yan14}. 
However, the author of Ref.$\,$\cite{Yan14} considered the expansion of $L(t)$ with respect to the exponential functions, which is not compatible with our method as discussed in Sec.$\,$\ref{num-bessel}.

We show in \ref{bk expansion} that one can extend Eq.$\,$(\ref{phi vs psi single}) to arbitrary phonon numbers.
It links the expansion functions $\psi_{j_1,\dots,j_K}^{(n)}$ to the total wave function, not only to the reduced density matrix.
For discrete baths, it was pointed out in Ref.$\,$\cite{LZBS14} that one can derive the total Wigner function from auxiliary density operators in the conventional approach. 
On the other hand, in our formalism, the total wave function can be obtained independent of the number of the $\{ a_i \}$-modes.

\subsection{Relevant degrees of freedom}
\label{link-relevant}

As shown in \ref{bk expansion}, the time evolution of the total wave function is obtained by solving that of $\psi_{j_1,\dots,j_K}^{(n)}$.
In terms of $\{ b_k \}$, the number of the modes is a finite number, $K$, even when that of $\{ a_i \}$-modes is infinite.
This fact implies that a large body of degrees of freedom in the $\{ a_i \}$ representation are actually irrelevant in the dynamics.

To understand this, one should first notice that the total wave function evolves in time as Eq.$\,$(\ref{te Psi}). 
This indicates that only those bath states are excited which are generated by acting $H_{\rm tot}$ to the initial state.
Now, let us assume that $H_{\rm tot}$ is given by Eq.$\,$(\ref{CL Hamiltonian}).
In this case, the operators applying to the bath state are $H_B$ and $X$.
To be specific, let us consider a contribution from the fourth order Taylor expansion of $\exp(-i H_{\rm tot} t / \hbar)$,
\begin{equation}
\label{ket - 1}
\ket{-} \equiv H_B X H_B X \ket{0}.
\end{equation}
Using the relation $H_B \ket{0} = 0$, one finds
\begin{equation}
\label{ket - 2}
\ket{-} =  X \left[ H_B , \left[ H_B, X \right] \right] \ket{0} + \left\{ \left[ H_B, X \right] \right\}^2 \ket{0}.
\end{equation}

As a simpler example, let us first investigate a case where all modes of phonon have the same energy, $H_B = \hbar \omega \sum_i a_i^{\dagger} a_i$. 
It was pointed out in Ref.$\,$\cite{KRNR93} and in Appendix C of Ref.$\,$\cite{HT12} that there is only one relevant mode in this case.
To show this, we first introduce $a = \sum_i d_i a_i / d$ with $d = \sqrt{\sum_i d_i^2}$, which satisfies the boson commutation relation $[a, a^{\dagger}] = 1$.
The commutation relation with $H_B$ is closed as $[H_B, a] = -\hbar \omega a$, and one finds that Eq.$\,$(\ref{ket - 2}) can be written only with $a$ and $a^{\dagger}$,
\begin{equation}
\label{ket - degenerate}
\ket{-} =  (\hbar \omega d)^2 (a + a^{\dagger})^2 \ket{0} + (\hbar \omega d)^2 (a-a^{\dagger})^2 \ket{0}.
\end{equation}
Therefore, the Hamiltonian operation as in Eq.$\,$(\ref{ket - 1})  generates only the bath states of the form $\left( a^{\dagger} \right)^{n} \ket{0}$.
This conclusion can be extended to arbitrary multiple operations of $H_B$ and $X$, and thus only $a$-mode is sufficient to describe the time evolution of the bath degrees of freedom.

When each mode is allowed to have different energies, on the other hand, the commutation relation with $H_B$ is not closed with respect to $a$, that is, $[H_B, a] = - \sum_i \hbar \omega_i d_i a_i$.
However, one can show that it is closed with respect to $\{ b_k \}$ and $\{ b_k^{\dagger} \}$. To see this, one should first notice
\begin{equation}
\label{H_B vs X a}
\begin{gathered}
\left[ H_B, X \right] = - \hbar \sum_i \omega_i d_i (a_i - a_i^{\dagger}), \\
\left[ H_B,  \left[ H_B, X \right] \right] = \hbar^2 \sum_i \omega_i^2 d_i (a_i + a_i^{\dagger}).
\end{gathered}
\end{equation}
From Eq.$\,$(\ref{b vs a sum}), one finds Eq.$\,$(\ref{X decomposition}) and
\begin{equation}
\label{omega a}
\begin{gathered}
\sum_{i} \omega_i d_i a_i^{\dagger} = i \hbar \sum_{k} \frac{dv_k}{dt} (0) b_k^{\dagger}, \\
\sum_{i} \omega_i^2 d_i a_i^{\dagger} = - \hbar \sum_{k} \frac{d^2 v_k}{dt^2} (0) b_k^{\dagger}.
\end{gathered}
\end{equation}
Combining Eqs.$\,$(\ref{H_B vs X a}) and (\ref{omega a}) gives
\begin{equation}
\label{H_B vs X b}
\begin{gathered}
\left[ H_B, X \right] = i \hbar^2 \sum_{k=1}^K \left\{ \frac{dv_k^*}{dt} (0) b_k + \frac{dv_k}{dt} (0) b_k^{\dagger} \right\}, \\
\left[ H_B,  \left[ H_B, X \right] \right] = - \hbar^3 \sum_{k=1}^K \left\{ \frac{d^2v_k^*}{dt^2} (0) b_k +  \frac{d^2v_k}{dt^2} (0) b_k^{\dagger} \right\}.
\end{gathered}
\end{equation}
Since $\{ b_k \}$ and $\{ b_k^{\dagger} \}$ satisfy the boson commutation relation, Eq.$\,$(\ref{b commutation}), all the bath states generated by the Hamiltonian operation as in Eq.$\,$(\ref{ket - 1}) can be described by mutual creation of the boson associated with the $\{ b_k^{\dagger} \}$ operators.
One can extend this conclusion to arbitrary multiple operations of $H_B$ and $X$. 
This explains why the finite $\{ b_k \}$-modes can be the only relevant degrees of freedom of the bath, even when the number of $\{ a_i \}$-modes is infinite. 


\section{Conclusions and future perspectives}
\label{conc}

For open quantum systems with a harmonic oscillator bath, we have developed a new method which is based on the phonon number representation of the bath degrees of freedom.
It enables one to calculate the reduced density matrix by following the time evolution of vectors rather than matrices.
This reduction is especially beneficial when the system of interest has a large dimension.
To formulate the method, we have extended the ideas behind the HEOM approach.
A numerical application to a quantum damped harmonic oscillator has shown that our method can well reproduce the exact results with a parameter set in which the damping of the amplitude is observed.
We have also shown that our method is particularly efficient for a problem with a weak and an intermediate coupling strength where the number of phonon to be excited is not so large, and for a non-Markovian case where the time scale of the bath is comparable with that of the system.
In our method, one can decompose the reduced density matrix into contribution of each phonon number.
It enables one to extract not only information on the degrees of freedom of the system, but also on the bath degrees of freedom such as the number of phonons to be excited in the course of time evolution.

We have also discussed a link of the hierarchy elements to the total wave function.
This consideration has naturally led us to an introduction of ladder operators, which satisfy the boson commutation relation.
In terms of the ladder operators, the explicit form of the total wave function has been derived. 
Furthermore, we have shown that the relevant degrees of freedom for the bath are finite, even when the number of harmonic oscillator modes is infinite.

In this paper, we have formulated the method based on the influence functional approach. 
As has been shown, this serves as a bottom-up approach. 
On the other hand, if one starts with the ladder operators $b_k^{\dagger} = d_i \bar{\eta}_k(\omega_i) a_i^{\dagger} / \hbar \ \ (k = 1,\dots,K)$, it is possible to reformulate the method as the coupled-channels approach. This is done in \ref{form ladder}. This serves as a top-down approach to the method.

There are several possible applications of the method proposed in this paper. 
One obvious way is an application to a fermionic bath. 
In this paper, we have considered a bosonic bath in which the ladder operators $\{ a_i \}$ satisfy the boson commutation relation $[a_i, a_j^{\dagger}] = \delta_{i,j}$. 
For fermionic baths, on the other hand, the ladder operators satisfy the fermion anti-commutation relation $\{ a_i, a_j^{\dagger} \} = \delta_{i,j}$. 
Such problems have been considered by several authors in order to investigate a dependence of the path to the equilibrium state on the statistics of the bath \cite{SAAL14}, and to treat an electronic transport phenomenon by explicitly taking into account the presence of electrode \cite{HC18,JZY08}.
We anticipate that similar discussions to Sec.$\,$\ref{link} is possible for these problems.

Not only the reduced density matrix, we have shown that the total wave function can also be extracted in our method.
The ML-MCTDH method, which is another powerful method to explore open quantum systems, derives the time evolution of the total wave function \cite{WT08}.
In the ML-MCTDH method, one takes into account each $\{ a_i \}$-degree of freedom explicitly.
However, it has been shown in Sec.$\,$\ref{link-relevant} that the $\{ b_k \}$-modes are the only relevant degrees of freedom in the Caldeiral-Leggett model.
The number of the $\{ a_i \}$-modes taken into account in the ML-MCTDH method amounts to hundreds to thousands. 
It might be possible to reduce the number by considering the $\{ b_k \}$ representation. 
If it is easy to increase the number of $\{ b_k \}$, $K$, it would provide a way to calculate long time behaviors (see discussions in Sec.$\,$\ref{num-bessel}).

Another future perspective is to apply our method to barrier transmission problems.
As can be seen from the formula of the reduced density matrix, Eq.$\,$(\ref{rhoS time evolution phonon}), our method enables one to calculate the two-time reduced density matrix defined by $\rho_{S}(q_a t_a; q_b t_b) \equiv {\rm Tr}_B \braket{q_a|\Psi (t_a)} \braket{\Psi (t_b)|q_b}$, which is an important quantity when one considers barrier transmission problems \cite{BT85}.
Notice that this quantity is very time consuming to calculate with the conventional HEOM approach, since one needs to scan the two-dimensional time space.
In marked contrast, our method makes it possible to calculate it with the same numerical cost as the calculation of the (single-time) reduced density matrix. 
Such work would serve to enlarge our basic understanding of quantum tunneling in a dissipative system.
A work towards this direction is now in progress, and we will report it in a separate publication.


\section*{Acknowledgments}
We would like to thank Dr. Denis Lacroix and Dr. Guillaume Hupin for their hospitality during a stay of M.T. at the IPN Orsay and for fruitful discussions in the early stage of this work.
We also thank Dr. Shimpei Endo for drawing our attention to the HEOM method.
We are grateful to Dr. Yusuke Tanimura for his suggestions regarding numerical calculations and figures. 
This work was supported by Tohoku University Graduate Program on Physics for
the Universe (GP-PU), and JSPS KAKENHI Grant Numbers JP18J20565 and 19K03861.


\appendix

\section{Derivation of the HEOM}
\label{derive HEOM}

In this appendix, we derive the HEOM, Eq.$\,$(\ref{HEOM}).
In the definition of the expansion functions Eq.$\,$(\ref{ef}), the time differentiation acts to three components on the right hand side, $\exp(i S_S[Q,t] / \hbar)$, $f[Q,t]$, and $Y_{j_1,\dots,j_K} [Q,t] \equiv \prod_{k=1}^K \{ y_k[Q,t] \}^{j_k}/i^{j_k} \sqrt{j_k !}$. 
As one sees in the derivation of the Schr\"odinger equation in the path integral formalism, the time derivative of the action term, $\exp(i S_S[Q,t] / \hbar)$, gives the Hamiltonian for the system $H_S$.

Using the representation of $f[Q,t]$ given by Eq.$\,$(\ref{f with v}), its time derivative reads
\begin{equation}
\frac{\partial f}{\partial t} [Q,t] = - h(Q(t)) \sum_{k=1}^{K} \bar{c}_k y_k[Q,t] f[Q,t]. 
\end{equation}
Note that inside the path integral $\int^{(q_a,t)}_{(q_c,0)} D[Q]$ in Eq.$\,$(\ref{ef}), $h(Q(t))$ turns into $h(q_a)$.

The time derivative of $y_k[Q,t]$ reads
\begin{equation}
\frac{\partial y_k}{\partial t} [Q,t] = \sum_{k^{\prime}=1}^{K} \bar{C}_{k, k^{\prime}} y_{k^{\prime}} [Q,t] + h(Q(t)) v_k(0).
\end{equation}
Here, we have used the fact that a set $\{ v_k \}$ is closed under differentiation (see Eq.$\,$(\ref{v derivative})).
From this equation, one finds
\begin{equation}
\begin{gathered}
\frac{\partial Y_{j_1,\dots,j_K}}{\partial t} [Q,t] = \sum_{k=1}^K j_k \bar{C}_{k, k} Y_{j_1,\dots,j_K} [Q,t] \\
+ \sum_{k \neq k^{\prime} = 1}^{K} \sqrt{j_k(j_{k^{\prime}}+1)} \ \bar{C}_{k, k^{\prime}} Y_{j_1,\dots,j_k-1,\dots,j_{k^{\prime}}+1,\dots,j_K}[Q,t] \\
- i h(Q(t)) \sum_{k=1}^K \sqrt{j_k} \ v_k(0) Y_{j_1,\dots,j_k-1,\dots,j_K}[Q,t].
\end{gathered}
\end{equation}
Combining all of these together, one finally obtains Eq.$\,$(\ref{HEOM}).

\section{The $\{ b_k \}$-modes representation of the total wave function}
\label{bk expansion}

In this appendix, we show that the expansion of the total wave function with respect to $\{ a_i^{\dagger} \}$ is equivalent to that with $\{ b_k^{\dagger} \}$,
\begin{equation}
\label{total wf expansion}
\begin{gathered}
\sum_{n=0}^{\infty} \sum_{(n_1+n_2+\dots=n)} \phi_{n_1,n_2,\dots}^{(n)} (q_a,t) \prod_{i} \frac{\left( a_i^{\dagger} \right)^{n_i}}{\sqrt{n_i !}} \ket{0} \\
= \sum_{n=0}^{\infty} \sum_{(j_1+ \dots + j_K = n)} \psi_{j_1,\dots,j_K}^{(n)}(q_a,t) \prod_{k=1}^{K} \frac{\left( b_k^{\dagger} \right)^{j_k}}{\sqrt{j_k !}} \ket{0}.
\end{gathered}
\end{equation}

From the definition of $\phi_{n_1,n_2,\dots}^{(n)} (q_a,t)$, Eq.$\,$(\ref{ef discrete}), the left hand side of Eq.$\,$(\ref{total wf expansion}) reads
\begin{equation}
\begin{gathered}
\int dq_c \ \varphi(q_c) \int_{(q_c,0)}^{(q_a,t)} D[Q] \, e^{i S_S[Q,t] / \hbar} f[Q,t] \\
\times \sum_{n=0}^{\infty} \sum_{(n_1+n_2+\dots=n)} \prod_{i} \frac{1}{n_i !} \left\{ \frac{1}{i} z_i[Q,t] a_i^{\dagger} \right\}^{n_i}.
\end{gathered}
\end{equation}
Using the relation $\sum_{i} z_i[Q,t] a_i^{\dagger} = \sum_{k=1}^{K} y_k[Q,t] b_k^{\dagger}$, which is obtained from Eq.$\,$(\ref{b vs a sum}), one finds
\begin{equation}
\begin{gathered}
\sum_{n=0}^{\infty} \sum_{(n_1+n_2+\dots=n)} \prod_{i} \frac{1}{n_i !} \left\{ \frac{1}{i} z_i[Q,t] a_i^{\dagger} \right\}^{n_i} \\
= \sum_{n=0}^{\infty} \frac{1}{n!} \left\{ \frac{1}{i} \sum_i z_i[Q,t] a_i^{\dagger} \right\}^{n} \\
= \sum_{n=0}^{\infty} \frac{1}{n!} \left\{ \frac{1}{i} \sum_{k=1}^{K} y_k[Q,t] b_k^{\dagger} \right\}^{n} \\
= \sum_{n=0}^{\infty} \sum_{(j_1 + \dots + j_K = n)} \prod_{k=1}^{K} \frac{1}{j_k !} \left\{ \frac{1}{i} y_k[Q,t] b_k^{\dagger} \right\}^{j_k},
\end{gathered}
\end{equation}
leading to Eq.$\,$(\ref{total wf expansion}).

Note that Eq.$\,$(\ref{total wf expansion}) does not imply that the bath states generated by $\prod_{k=1}^{K} (b_k^{\dagger})^{j_k} / \sqrt{j_k !} \ket{0}$ are a complete set in the bath space. 
They are orthogonal to each other (see Eq.$\,$(\ref{j1jK base orthogonal})), and simply span a subspace of the total bath space.  
As long as one considers the Caldeira-Leggett model, Eq.$\,$(\ref{CL Hamiltonian}), however, it is sufficient to work on such subspace (see Sec.$\,$\ref{link-relevant}).

\section{Formulation based on the ladder operators}
\label{form ladder}

In this appendix, we reformulate the method presented in this paper based only on the $\{ b_k \}$ and $\{ b_k^{\dagger} \}$ algebra, without referring to the influence functional method. 
To be more specific, we derive the HEOM, Eq.$\,$(\ref{HEOM}), and the formula for the reduced density matrix, Eq.$\,$(\ref{rhoS time evolution phonon}).

We first introduce the eigenstates of the $\{ b_k \}$-modes as
\begin{equation}
\ket{j_1, \dots, j_K} \equiv \prod_{k=1}^{K} \frac{\left( b_k^{\dagger} \right)^{j_k}}{\sqrt{j_k !}} \ket{0}.
\end{equation}
From the commutation relation between $\{ b_k \}$ and $\{ b_k^{\dagger} \}$, Eq.$\,$(\ref{b commutation}), one finds the orthogonal relation,
\begin{equation}
\label{j1jK base orthogonal}
\braket{j_1,\dots,j_K|l_1,\dots,l_K} = \prod_{k=1}^K \lambda_k^{j_k} \delta_{j_k, l_k}.
\end{equation}
Using this basis, we expand the total wave function $\ket{\Psi (t)}$, as 
\begin{equation}
\label{j1jK base expansion}
\braket{q_a|\Psi (t)} = \sum_{n=0}^{\infty} \sum_{(l_1+\dots+l_K = n)} \psi_{l_1,\dots,l_K}^{(n)} (q_a,t) \ket{l_1,\dots,l_K}.
\end{equation}
Although this expansion does not span the whole bath space, it is sufficient for the Caldeira-Leggett model as has been discussed in \ref{bk expansion}.

Let us first derive the HEOM.
The total wave function $\ket{\Psi (t)}$ satisfies the Schr\"odinger equation with the total Hamiltonian given by Eq.$\,$(\ref{CL Hamiltonian}). Taking an inner product with $\bra{q_a,j_1,\dots,j_K} \equiv \bra{q_a}\bra{j_1,\dots,j_K}$ gives
\begin{equation}
\label{sch total}
\braket{q_a,j_1,\dots,j_K| i \hbar \frac{\partial}{\partial t} |\Psi (t)} = \braket{q_a,j_1,\dots,j_K| H_{\rm tot} |\Psi (t)}.
\end{equation}
In what follows, we show that this leads to the HEOM, Eq.$\,$(\ref{HEOM}). 
For convenience, we set $j_1 + \dots + j_K = n$ and introduce $\Lambda \equiv \prod_{k=1}^K \lambda_k^{j_k}$.

Since the time derivative and $H_S$ do not act on to the bath degrees of freedom, those contributions in Eq.$\,$(\ref{sch total}) are given by
\begin{equation}
\label{sch dt}
\braket{q_a,j_1,\dots,j_K| i \hbar \frac{\partial}{\partial t} |\Psi (t)} = \Lambda i \hbar \frac{\partial}{\partial t} \psi_{j_1,\dots,j_K}^{(n)} (q_a,t),
\end{equation}
and
\begin{equation}
\label{sch hs}
\braket{q_a,j_1,\dots,j_K| H_S |\Psi (t)} = \Lambda H_S(q_a) \psi_{j_1,\dots,j_K}^{(n)} (q_a,t),
\end{equation}
respectively.

To compute the contribution of  $H_B$ in Eq.$\,$(\ref{sch total}), one needs to estimate the quantity
\begin{equation}
\braket{j_1,\dots,j_K| a_i^{\dagger} a_i |l_1,\dots,l_K}.
\end{equation}
This can be carried out with the Wick's theorem \cite{FW71}.
$a_i$ is contracted with one of $\ket{l_1,\dots,l_K}$. 
The contraction with $b_{k^{\prime}}^{\dagger}$ has $l_{k^{\prime}}$ choices, and is given by $d_i \bar{\eta}_{k^{\prime}}(\omega_i) / \hbar$. 
The remaining bath state becomes $\ket{l_1,\dots,l_{k^{\prime}}-1,\dots,l_K}/\sqrt{l_k^{\prime}}$. 
One can do the same for the contraction of $a_i^{\dagger}$. 
These considerations lead to
\begin{equation}
\begin{gathered}
\braket{j_1,\dots,j_K| a_i^{\dagger} a_i |l_1,\dots,l_K} = \frac{d_i^2}{\hbar^2} \sum_{k,k^{\prime} = 1}^K  \bar{\eta}_{k^{\prime}} (\omega_i) \bar{\eta}_k^* (\omega_i) \sqrt{j_k l_{k^{\prime}}} \\
\times \frac{\Lambda}{\lambda_k} \delta_{j_1, l_1} \dots \delta_{j_k-1, l_k} \dots \delta_{j_{k^{\prime}}, l_{k^{\prime}}-1} \dots \delta_{j_K, l_K}.
\end{gathered}
\end{equation}
Hence, one finds
\begin{equation}
\label{sch hb 0}
\begin{gathered}
\braket{q_a,j_1,\dots,j_K| H_B |\Psi (t)} \\
= \Lambda \sum_{k=1}^K j_k \frac{1}{\lambda_k} \left\{ \sum_i \hbar \omega_i \frac{d_i^2}{\hbar^2} \bar{\eta}_{k} (\omega_i) \bar{\eta}_k^* (\omega_i) \right\} \psi_{j_1,\dots,j_K}^{(n)} (q_a,t) \\
+ \Lambda \sum_{k \neq k^{\prime}=1}^K \sqrt{j_k (j_{k^{\prime}}+1)} \frac{1}{\lambda_k} \left\{ \sum_i \hbar \omega_i \frac{d_i^2}{\hbar^2} \bar{\eta}_{k^{\prime}} (\omega_i) \bar{\eta}_k^* (\omega_i) \right\} \\
\times \psi_{j_1,\dots,j_k-1,\dots,j_{k^{\prime}}+1,\dots,j_K}^{(n)} (q_a,t).
\end{gathered}
\end{equation}
Notice that the time derivative of $L(t)$ can be represented in several ways,
\begin{equation}
\begin{gathered}
\frac{1}{\hbar} \frac{d}{dt_1} L(t_1-t_2) =  -i \sum_i \omega_i \frac{d_i^2}{\hbar^2} e^{-i\omega_i t_1} e^{i \omega_i t_2} \\
= - i \sum_{k,k^{\prime}=1}^K \left\{ \sum_i \omega_i \frac{d_i^2}{\hbar^2} \bar{\eta}_{k^{\prime}} (\omega_i) \bar{\eta}_k^* (\omega_i)  \right\} v_{k^{\prime}}(t_1) v_{k}^*(t_2) \\
= \sum_{k,k^{\prime}=1}^K \lambda_k \bar{C}_{k, k^{\prime}} v_{k^{\prime}}(t_1) v_{k}^*(t_2),
\end{gathered}
\end{equation}
where the last two equations suggest
\begin{equation}
\sum_i \hbar \omega_i \frac{d_i^2}{\hbar^2} \bar{\eta}_{k^{\prime}} (\omega_i) \bar{\eta}_k^* (\omega_i) = i \hbar \lambda_k \bar{C}_{k, k^{\prime}}.
\end{equation}
Substituting this into Eq.$\,$(\ref{sch hb 0}), one obtains
\begin{equation}
\label{sch hb}
\begin{gathered}
\braket{q_a,j_1,\dots,j_K| H_B |\Psi (t)} \\
= \Lambda i \hbar \sum_{k=1}^K j_k \ \bar{C}_{k, k} \psi_{j_1,\dots,j_K}^{(n)} (q_a,t) \\
+ \Lambda i \hbar \sum_{k \neq k^{\prime}=1}^K \sqrt{j_k (j_{k^{\prime}}+1)} \ \bar{C}_{k, k^{\prime}} \psi_{j_1,\dots,j_k-1,\dots,j_{k^{\prime}}+1,\dots,j_K}^{(n)} (q_a,t).
\end{gathered}
\end{equation}

Finally, since the interaction Hamiltonian can be represented as $H_I = \hbar h(q) \sum_{k=1}^K ( v_k^{*}(0) b_k + v_k(0) b_k^{\dagger})$, the contribution of this term in Eq.$\,$(\ref{sch total}) reads
\begin{equation}
\label{sch hi}
\begin{gathered}
\braket{q_a,j_1,\dots,j_K| H_I |\Psi (t)} \\
= \Lambda h(q_a) \sum_{k=1}^K \sqrt{j_k + 1} \ \hbar \bar{c_k} \psi_{j_1,\dots,j_k+1,\dots,j_K}^{(n+1)}(q_a,t) \\
+ \Lambda h(q_a) \sum_{k=1}^K \sqrt{j_k} \ \hbar v_k(0) \psi_{j_1,\dots,j_k-1,\dots,j_K}^{(n-1)}(q_a,t),
\end{gathered}
\end{equation}
where we have used $\bar{c}_k = \lambda_k v_k^{*}(0)$.
Substituting Eqs.$\,$(\ref{sch dt}), (\ref{sch hs}), (\ref{sch hb}), and (\ref{sch hi}) into Eq.$\,$(\ref{sch total}), and dividing it by $\Lambda$, one obtains Eq.$\,$(\ref{HEOM}).

Next, we derive the formula for the reduced density matrix.
Regarding the expansion of the total wave function given by Eq.$\,$(\ref{j1jK base expansion}), 
one simply needs to take a partial trace over the subspace spanned by the basis $\{ \ket{j_1,\dots,j_K} \}$, which reads
\begin{equation}
\sum_{j_1,\dots,j_K = 0}^{\infty} \frac{\ket{j_1,\dots,j_K}\bra{j_1,\dots,j_K}}{\prod_{k=1}^K \lambda_k^{j_k}}.
\end{equation}
With this procedure, one obtains Eq.$\,$(\ref{rhoS time evolution phonon}).

The same formula is derived with the trace with $\{ \ket{n_1,n_2,\dots} \}$.
To show this, one should first notice 
\begin{equation}
\begin{gathered}
\sum_{(n_1+n_2+\dots=n)} \ket{n_1,n_2,\dots}\bra{n_1,n_2,\dots} \\
= \frac{1}{n !} \sum_{i_1,\dots,i_n} a_{i_1}^{\dagger} \dots a_{i_n}^{\dagger} \ket{0}\bra{0} a_{i_1} \dots a_{i_n}.
\end{gathered}
\end{equation}
Using this representation, the reduced density matrix is given by
\begin{equation}
\label{j1jK rhoS}
\begin{gathered}
\rho_S(q_a,q_b,t) = {\rm Tr}_B \braket{q_a|\Psi (t)} \braket{\Psi (t)|q_b} \\
= \sum_{n=0}^{\infty} \frac{1}{n!} \sum_{(j_1+\dots+j_K = n)} \sum_{(l_1+\dots+l_K = n)} \psi_{j_1,\dots,j_K}^{(n)} (q_a,t) \left\{ \psi_{l_1,\dots,l_K}^{(n)} (q_b,t) \right\}^* \\
\times \sum_{i_1,\dots,i_n} \braket{0| a_{i_1} \dots a_{i_n} |j_1,\dots,j_K} \braket{l_1,\dots,l_K| a_{i_1}^{\dagger} \dots a_{i_n}^{\dagger} |0}.
\end{gathered}
\end{equation}
To evaluate the matrix elements, the Wick's theorem can be utilized. 
Because of Eq.$\,$(\ref{eta vs lambda}), when $a_{i_1}$ is contracted with $b_{k_1}^{\dagger}$, $a_{i_1}^{\dagger}$ should be contracted with the same mode, $b_{k_1}$. This indicates $l_k = j_k$ for $k = 1,\dots,K$. There are $j_{k_1}^2$ such choices. 
Repeating this procedure to $a_{i_n}$, one finds
\begin{equation}
\begin{gathered}
\sum_{i_1,\dots,i_n} \braket{0| a_{i_1} \dots a_{i_n} |j_1,\dots,j_K} \braket{j_1,\dots,j_K| a_{i_1}^{\dagger} \dots a_{i_n}^{\dagger} |0} \\
= \prod_{k=1}^K j_k ! \sum_{(k_1,\dots,k_n) = (j_1,\dots,j_K)} \lambda_{k_1} \dots \lambda_{k_n} ,
\end{gathered}
\end{equation}
where $\sum_{(k_1,\dots,k_n) = (j_1,\dots,j_K)}$ means a sum over $k_i = 1,\dots,K$ for $i = 1,\dots,n$, with a constraint such that $q$ appears $j_q$ times for $q = 1,\dots,K$. 
Since the number of such combinations is $n!/\prod_{k=1}^K j_k!$, the matrix elements read
\begin{equation}
\sum_{i_1,\dots,i_n} \braket{0| a_{i_1} \dots a_{i_n} |j_1,\dots,j_K} \braket{j_1,\dots,j_K| a_{i_1}^{\dagger} \dots a_{i_n}^{\dagger} |0} = n! \prod_{k=1}^K \lambda_k^{j_k}.
\end{equation}
Substituting this into Eq.$\,$(\ref{j1jK rhoS}), one finally obtains Eq.$\,$(\ref{rhoS time evolution phonon}).


\end{document}